\documentclass{nature}
\usepackage{amsmath,amssymb,graphicx,hyperref,url}  
\usepackage{ulem}
\usepackage{lineno}
\usepackage[none]{hyphenat}
\usepackage[usenames]{xcolor}
\usepackage{adjustbox}

\newcommand{\tc}{\textcolor{black}}
\newcommand{\tx}{\textcolor{black}}
\newcommand{\tp}{\textcolor{black}}

\def\etal{{\frenchspacing\it et al.}}
\def\ie{{\frenchspacing\it i.e.}}
\def\eg{{\frenchspacing\it e.g.}}

\def\be{\begin{equation}}
\def\ee{\end{equation}}
\def\ba{\begin{eqnarray}}
\def\ea{\end{eqnarray}}

\title{Extracting high-order cosmological information in galaxy surveys with power spectra}

\author{Yuting Wang$^{1,2}$,
Gong-Bo Zhao$^{1,2,3}$ \thanks{\url{email: gbzhao@nao.cas.cn}}, 
Kazuya Koyama$^{4}$ \thanks{\url{email: kazuya.koyama@port.ac.uk}},
Will J. Percival$^{5,6,7}$ \thanks{\url{email: will.percival@uwaterloo.ca}},
Ryuichi Takahashi$^{8}$,
Chiaki Hikage$^{9}$,
H\'ector Gil-Mar\'{\i}n$^{10}$,
ChangHoon Hahn$^{11}$,
Ruiyang Zhao$^{1,3}$,
Weibing Zhang$^{1,3}$,
Xiaoyong Mu$^{1,3}$,
Yu Yu$^{12}$,
Hong-Ming Zhu$^{13}$,
Fei Ge$^{14}$ \\
$^{1}$ National Astronomy Observatories, Chinese Academy of Sciences, Beijing, 100101, P.R.China \\
$^{2}$ Institute for Frontiers in Astronomy and Astrophysics, Beijing Normal University, Beijing 102206, China\\
$^{3}$ School of Astronomy and Space Science, University of Chinese Academy of Sciences, Beijing 100049, P.R.China\\
$^{4}$ Institute of Cosmology and Gravitation, University of Portsmouth, Dennis Sciama Building, Portsmouth PO1 3FX, United Kingdom\\
$^{5}$ Waterloo Centre for Astrophysics, University of Waterloo, 200 University Ave W, Waterloo, ON, N2L 3G1, Canada\\
$^{6}$ Department of Physics and Astronomy, University of Waterloo, 200 University Ave W, Waterloo, ON, N2L 3G1, Canada\\
$^{7}$ Perimeter Institute for Theoretical Physics, 31 Caroline St. North, Waterloo, ON, N2L 2Y5, Canada\\
$^{8}$ Faculty of Science and Technology, Hirosaki University, 3 Bunkyo-cho, Hirosaki, Aomori 036-8561, Japan\\
$^{9}$ Kavli Institute for the Physics and Mathematics of the Universe (Kavli IPMU, WPI), University of Tokyo, 5-1-5 Kashiwanoha, Kashiwa, Chiba, 277-8583, Japan\\
$^{10}$ ICC, University of Barcelona, IEEC-UB, Mart\'{\i} i Franqu\`es, 1, E-08028 Barcelona, Spain\\
$^{11}$ Department of Astrophysical Sciences, Princeton University, Peyton Hall, Princeton NJ 08544, USA\\
$^{12}$ Department of Astronomy, Shanghai Jiao Tong University, Shanghai, 200240, P.R.China\\
$^{13}$ Canadian Institute for Theoretical Astrophysics, University of Toronto, 60 St. George Street, Toronto, Ontario M5S 3H8, Canada\\
$^{14}$ Department of Physics and Astronomy, University of California, Davis, California 95616, USA
}

\begin{document}
\maketitle

\vspace{1cm}

\begin{abstract}
\section{Abstract}
 The reconstruction method was proposed more than a decade ago to boost the signal of baryonic acoustic oscillations measured in galaxy redshift surveys, which is one of key probes for dark energy. After moving the observed overdensities in galaxy surveys back to their initial position, the reconstructed density field is closer to a linear Gaussian field, with higher-order information moved back into the power spectrum. We find that by jointly analysing power spectra measured from the pre- and post-reconstructed galaxy samples, higher-order information beyond the $2$-point power spectrum can be efficiently extracted, which generally yields an information gain upon the analysis using the pre- or post-reconstructed galaxy sample alone. This opens a window to easily use higher-order information when constraining cosmological models. 

\end{abstract}

\section{Introduction}

The science driver for massive galaxy spectroscopic surveys is to extract cosmological information from the clustering of galaxies in the past lightcone. Baryon acoustic oscillations (BAO)\cite{BAO98}, formed in the early Universe due to interactions between photons and baryons under pressure and gravity, yield a special clustering pattern of galaxies around a characteristic comoving scale around $150$ Mpc, which is one of key probes for dark energy\cite{DE1,DE2}. The increasing size of galaxy redshift surveys over the decade $2000$-$2010$ led ultimately to a $5\sigma$ detection of BAO by the Baryon Oscillation Spectroscopic Survey (BOSS)\cite{BOSS}. This enabled the BAO to be used as an accurate {\it standard ruler} to measure the geometry of the Universe and constrain the cosmic expansion history. A compilation of results from the Sloan Digital Sky Survey (SDSS) galaxy survey recently demonstrated the power of this technique\cite{eBOSSDR16}.

The BAO feature is generally blurred by the nonlinear evolution of the Universe reducing its strength as a standard ruler, and various reconstruction methods have been developed to sharpen the BAO peak by undoing the nonlinear evolution of the density field. The commonly used Lagrangian reconstruction, for example, linearises the density field by shifting the galaxies using the displacement field \cite{recon07,recon12,reconLag}, while for the Eulerian reconstruction, manipulation is performed at the field level without moving the galaxies\cite{reconEuler}. 

Although designed to boost the BAO signal originally, the reconstruction method can, in principle, also better extract the general cosmological information from the clustering. For example, the redshift space distortions (RSD)\cite{Kaiser,RSD1,RSD2}, which is caused by peculiar motions of galaxies under gravity, can also be better constrained using the reconstructed sample\cite{Hikage2020}.

{The standard method for BAO-reconstruction alters the over-density field so that it is more correlated with the initial linear field\cite{Zhu18}, and the level of mode-coupling can be highly reduced\cite{Seo15}.} It achieves this by inferring the bulk-flows using the observed galaxy field and then removing these displacements from both the galaxy positions and the map of expected density setting the baseline from which the over-densities are found. The power spectrum of the post-reconstructed sample ($P_{\rm post}$) provides additional information for cosmology compared with the pre-reconstructed sample ($P_{\rm pre}$) because the reconstruction restores the linear signal reduced by the non-linear evolution. 

The higher order statistics such as $B$, the bispectrum, induced by the non-linear evolution are, in turn, reduced. Thus we can extract more cosmological information encoded in the linear density field from $P_{\rm post}$ than $P_{\rm pre}$. On the other hand, the non-linear field contains information on small-scale clustering such as galaxy biases, { which can provide} better constraints on cosmological parameters by breaking the degeneracy between them. For the pre-reconstructed sample, this information can be extracted by combining the power spectrum with higher-order statistics. However, the power spectrum and higher order statistics such as the bispectrum are correlated, reducing our ability to estimate cosmological parameters. If we instead consider the post-reconstructed power spectrum, the covariance between the power spectrum and the bispectrum is reduced, and we can extract the information more efficiently. In this work, we show that the same improvement can be achieved by a joint analysis of $P_{\rm pre}$, $P_{\rm post}$ and $P_{\rm cross}$ (the cross-power spectrum between the pre- and post-reconstructed density fields). Due to the restored linear signal in the reconstructed density field, $P_{\rm post}$ is decorrelated with $P_{\rm pre}$ on small scales, which are dominated by the non-linear { information}. On these scales, the combination of $P_{\rm pre}$, $P_{\rm post}$ and $P_{\rm cross}$ has a similar ability to extract cosmological information as the combination of $P_{\rm post}$ or $P_{\rm pre}$ with the bispectrum because we are able to use the linear information in $P_{\rm post}$ and higher-order information in $P_{\rm pre}$ separately. 

Let us rewrite the non-linear over-density field as $\delta = R+\Delta$, where $R$ is the over-density field after reconstruction, which is closer to the linear field. It is then straightforward to express $P_{\rm pre}$, $P_{\rm cross}$ in terms of $P_{RR}(=P_{\rm post})$, $P_{\Delta \Delta}$ (the power spectrum of $\Delta$) and $P_{\Delta R}$ (the cross-power spectrum between $\Delta$ and $R$). Using perturbation theory\cite{Hikage19}, we can show that, at the leading order, $P_{\Delta \Delta}$ contains the integrated contribution from the bispectrum of squeezed-limit triangles while $P_{R \Delta}$ contains the integrated contribution from the trispectrum (T) of folded/squeeze-limit quadrilaterals (See Supplementary Note 1 for an explanation). In this fashion when combining $P_{\rm pre}$ and $P_{\rm cross}$ with $P_{\rm post}$, we are essentially adding in higher-order signal, thus naturally gaining information. Note that in order to match the information obtained by adding these two extra statistics, it is not enough to consider the bispectrum signal of the pre-reconstructed field, but both bispectrum and trispectrum signals. For this reason the information content of $P_{\rm pre}+B$ is different from that contained in $P_{\rm pre}+P_{\rm cross}+P_{\rm post}$. However, it is important to note that higher-order information that reconstruction brings is only a part from the total contained in the full bispectrum and trispectrum data-vectors. This is why a full analysis using $P+B+T$ will always provide more information. However, such an analysis is not very practical because of the size of the full data-vector and the computational time typically required to measure $B$ and especially $T$ directly. In this paper we show that $P_{\rm pre}+P_{\rm cross}+P_{\rm post}$ is an efficient alternative for extracting the relevant information from higher-order statistics for cosmological analyses.

To demonstrate the power of jointly using density fields before and after the reconstruction, we perform an anisotropic Lagrangian reconstruction (see Methods for details) on each realisation of the {\sc Molino} galaxy mocks\cite{Molino}, which is a large suite of realistic galaxy mocks produced from the Quijote simulations\cite{Quijote} at $z=0$. We then use these mocks to calculate the data covariance matrix and derivatives numerically for a Fisher matrix analysis\cite{Fisher} using the measured multipoles (up to $\ell=4$) of $P_{\rm pre}$, $P_{\rm post}$ and $P_{\rm cross}$ on the parameter set $ \mathbf{\Theta} \equiv \{\Omega_{\rm m}, \Omega_{\rm b}, h, n_s, \sigma_8, M_{\nu}, \mathbf{H}\}$ where $\mathbf{H}$ denotes the Halo Occupation Distribution (HOD) parameters, \ie, $\mathbf{H}\equiv\{\log M_{\rm min}, \sigma_{\log M}, \log M_0, \alpha, \log M_1\}$\cite{Zheng07} (see Methods for details).

\section{Results and Discussion}

Panel {\bf a} in Figure 1 shows the measured power spectra monopole (the quadrupule and hexadecapole are shown in Supplement Figure 1), and we see that $P_{\rm cross}$ decreases dramatically with scale compared to $P_{\rm pre}$ and $P_{\rm post}$. This indicates a decorrelation between $P_{\rm pre}$ and $P_{\rm post}$ below quasi-nonlinear scales ($k\gtrsim0.1~h~{\rm Mpc}^{-1}$), which is largely due to the difference in infrared effects contained in density fluctuations before and after the BAO reconstruction\cite{Sugiyama2024}.

From the original data vector $\{P_{\rm pre}, P_{\rm post}, P_{\rm cross}\}$, we can construct their linear combinations, $P_{\rm R \Delta}, P_{\rm \Delta \Delta}, P_{\rm \delta \Delta}$ defined as 
\ba
P_{\rm R \Delta} = P_{\rm cross} - P_{\rm post} \ ; P_{\rm \Delta \Delta} = P_{\rm pre} + P_{\rm post} -2 P_{\rm cross} \ ; 
  P_{\rm \delta \Delta} = P_{\rm pre}-P_{\rm cross}. 
 \label{eq:rotated}
\ea
Panel {\bf b} in Figure 1 show these power spectra. As discussed above, these power spectra involving $\Delta$ contain the information {of} part of the high-order statistics such as bispectrum and trispectrum.

The derivatives of $\{P_{\rm pre}, P_{\rm post}, P_{\rm cross}\}$ with respect to cosmological parameters and HOD parameters are presented in Supplement Figures 2-4. We have checked and confirmed the convergence of our Fisher matrix result given the number of mocks available, demonstrating the robustness of our result (see Supplement Note 2 and Supplement Figure 5 for details). 

The correlation matrix for the monopole of power spectrum and bispectrum (only the correlation with the squeezed-limit of $B_0$ is visualised for brevity) is shown in Figure 2. It is seen that $P_0^{\rm pre}$ highly correlates with $B_0$, confirming that the bispectrum is induced by nonlinearities. In contrast, $P_0^{\rm post}$ weakly correlates with $B_0$, or with $P_0^{\rm pre}$ and $P_0^{\rm cross}$ on nonlinear scales (\eg, at $k\gtrsim0.2~h~{\rm Mpc}^{-1}$). This, however, does not mean that $P_{\rm post}$ is irrelevant to the bispectrum -- it actually is a mixture of $P_{\rm pre}$ and certain integrated forms of the bispectrum and trispectrum information\cite{Hikage19,reconEuler}. Therefore by combining $P_{\rm post}$ with $P_{\rm pre}$ and $P_{\rm cross}$, one can in principle decouple the leading contribution in the power spectrum, bispectrum and trispectrum. The integrated form of the bispectrum information dominates $P_{\Delta\Delta}$, which strongly correlates with $B_0$, as shown in Supplement Figure 6 (see Supplement Note 3). The fact that $P_{\rm post}$ barely correlates with $B_0$ implies that the information content in $P_{\rm post}$ combined with $B_0$ may be similar to that in $P_{\rm post}$ combined with $P_{\rm pre}$ and $P_{\rm cross}$, which is confirmed to be the case by the results from the Fisher analysis presented below.

The cumulative signal-to-noise ratio (SNR) for power spectrum multipoles (up to $\ell=4$), bispectrum monopole, and various data combinations are shown in Supplement Figure 7, in which we can see that the joint $2$-point statistics, $P_{\rm all}$, is measured with a greater SNR than that of $P_{\rm post}$ or $P+B_0$, which may mean that $P_{\rm all}$ can be more informative than $P_{\rm post}$ or $P+B_0$ for constraining cosmological parameters.

To confirm the constraining power of $P_{\rm all}$, we then project the information content in the observables onto cosmological parameters using a Fisher matrix approach. Contour plots for $({\rm log}M_{\rm 0}, \sigma_8)$ derived from different datasets with two choices of $k_{\rm max}$ (the maximal $k$ for the observables used in the analysis) are shown in Figure 3 (more complete contour plots are shown in Supplement Figures 8-12). The smoothing scale is set to be $S=10~h^{-1}~{\rm Mpc}$ when performing the reconstruction. The degeneracies between parameters using $P_{\rm pre}, P_{\rm post}$ and $P_{\rm cross}$ are generally different, because $P_{\rm pre}, P_{\rm post}$ and $P_{\rm cross}$ differ to a large extent in terms of nonlinearity on small scales. This is easier to see in Supplement Figure 8, in which contours for the same parameters are shown for observables used in several $k$ intervals. The contours derived from $P_{\rm pre}$ and $P_{\rm post}$ generally rotate as $k$ increases because of the kick-in of nonlinear effects, which affects $P_{\rm pre}$ and $P_{\rm post}$ at different levels on the same scale. This significantly improves the constraint when these power spectra are combined, labelled as $P_{\rm all}$, which is tighter than that from the traditional joint power spectrum-bispectrum analysis ($P_{\rm pre}+B_0$). It is found that $P_{\rm all}$ can even win against $P_{\rm post}+B_0$ in some cases, demonstrating the robustness of this method. The contour plots with 1D posterior distributions for all parameters with $S=10$ and $20~h^{-1}~{\rm Mpc}$ and $k_{\rm max}=0.2$ and $0.5~h~{\rm Mpc}^{-1}$ are shown in Supplement Figures 9-12, respectively. In all cases, $P_{\rm all}$ offers competitive constraints on all parameters, even compared to the joint $P_{\rm post}+B_0$ analysis.

To further quantify our results, in Figure 4 we compare the square root of the Fisher matrix element for each parameter, with and without marginalising over others, derived from $P_{\rm all}$ and $P_{\rm pre}+B_0$, respectively, with two choices of $k_{\rm max}$.

For $k_{\rm max}=0.2~h~{\rm Mpc}^{-1}$, we see that the Fisher information for each parameter (panel ${\textbf a}$: without marginalising over others) derived from $P_{\rm all}$ is identical or even greater than that in $P_{\rm pre}+B_0$. In other words, combining all power spectra we can efficiently extract the information in $P_{\rm pre}+B_0$. After marginalising over other parameters, panel ${\textbf b}$ shows that the uncertainty on each parameter gets redistributed due to the degeneracy. The ratios for the HOD parameters are all greater than unity especially for ${\rm log} M_0$ and $\sigma_{{\rm log}M}$, demonstrating the power of our method on constraining HOD parameters. The information content for cosmological parameters in $P_{\rm pre} +B_0$ is well recovered by using $P_{\rm all}$, although the recovery for $M_{\nu}$ is relatively worse. The overall trend for the case of $k_{\rm max}=0.5~h~{\rm Mpc}^{-1}$ is similar, although the advantage of using $P_{\rm all}$ over $P_{\rm pre} +B_0$ gets degraded to some extent. However, $P_{\rm all}$ is still competitive: it almost fully recovers the information for the HOD parameters in $P_{\rm pre}+B_0$ with or without marginalisation, and largely wins against $P_{\rm pre}+B_0$ after marginalisation. Regarding the cosmological parameters, $P_{\rm all}$ recovers all information in $P_{\rm pre} +B_0$ before the marginalisation, although the recovery is slightly worse for $M_{\nu}$. After marginalisation when the uncertainties are redistributed, the constraint from $P_{\rm all}$ is generally worse than $P_{\rm pre} +B_0$, especially for $M_{\nu}$.

The 68\% confidence level constraints on each parameter fitting to various datasets are shown in Table~1. To quantify the information gain, we evaluate the Figure-of-Merit (FoM) defined as $\left[{\bf det} (F)\right]^{1/(2N_{\rm p})}$, where $F$ denotes the Fisher matrix and $N_{\rm p}$ is the total number of free parameters. For the ease of comparison, for cases with different $k_{\rm max}$, we normalise all the quantities using the corresponding one for $P_{\rm pre}$. As shown, for $k_{\rm max}=0.2~h~{\rm Mpc}^{-1}$, $({\rm FoM})_{P_{\rm all}}$ is greater than all others, namely, it is larger than $({\rm FoM})_{P_{\rm pre}}$ and $({\rm FoM})_{P_{\rm post}}$ by a factor of $2.7$ and $1.7$, respectively, and it is even greater than $({\rm FoM})_{P_{\rm post}+B_0}$ by $\sim13\%$. For $k_{\rm max}=0.5~h~{\rm Mpc}^{-1}$, $({\rm FoM})_{P_{\rm all}}$ is also more informative than $({\rm FoM})_{P_{\rm pre}}$ and $({\rm FoM})_{P_{\rm post}}$ by a factor of $2.1$ and ${ 1.5}$, respectively, and is the same as $({\rm FoM})_{P_{\rm pre}+B_0}$, but is less than $({\rm FoM})_{P_{\rm post}+B_0}$ by $\sim10\%$ in this case.

To highlight the constraining power on cosmological parameters, we also list ${\rm FoM}_{\rm cos}$, which is the FoM with all HOD parameters fixed. It shows a similar trend as ${\rm FoM}_{\Theta}$: $P_{\rm all}$ is the most informative data combination for $k_{\rm max}=0.2~h~{\rm Mpc}^{-1}$, but it is outnumbered by ${P_{\rm pre}+B_0}$ and ${P_{\rm post}+B_0}$ by $13$\% and ${ 30\%}$, respectively, for the case of $k_{\rm max}=0.5~h~{\rm Mpc}^{-1}$.

\section{Conclusions}

As demonstrated in this analysis, a joint analysis using $P_{\rm pre}, P_{\rm post}$ and $P_{\rm cross}$ is an efficient way to extract high-order information from galaxy catalogues, and in some cases, $P_{\rm all}$ is more informative even than $P_{\rm post}+B_0$, which is computationally much more expensive.

In this example, the $k$-binning for $P$ and $B$ are different, namely, \tc{$\Delta k (B) = 3 k_f \sim 0.019 \ h {\rm Mpc}^{-1} \sim 1.9 \  \Delta k (P)$} where $k_f$ denotes the fundamental $k$ mode given the box size of the simulation. We have checked that using \tc{a finer $k$-binning for $B$ only improves the constraints marginally\cite{Binfo}, namely}, 
the FoM can only be raised by $\sim10\%$ when $\Delta k (B)$ is reduced from $3 k_f$ to $k_f$, which is largely due to the strong mode-coupling in $B$ as shown in Figure 2. Such a fine binning is not practical anyway as, for example, using $\Delta k (B) = k_f$ up to $k=0.5~h~{\rm Mpc}^{-1}$, we end up with more than $50,000$ data points to measure for $B_0$.

Note that the {\sc Molino} mock is produced at $z=0$, where the nonlinear effects are the strongest. At higher redshifts, the density fields are more linear and Gaussian, thus we may expect less gain from our method. This can be seen from panel {\bf a} of Supplement Figure 13, in which the correlation between $P_{\rm pre}$ and $P_{\rm post}$ at various redshifts is shown. As expected, $P_{\rm pre}$ and $P_{\rm post}$ are more correlated at higher redshifts, \eg, the correlation approaches $0.95$ at $z=5$ around $k\sim0.3 \ h \ {\rm Mpc}^{-1}$, which implies that almost no information gain can be obtained at such high redshifts. As argued previously, the decorrelation at lower $z$ is due to the fact that $P_{\rm pre}$ and $P_{\rm post}$ contain different levels of nonlinearity, as illustrated in panel {\bf b}, thus are complementary. Also, the Alcock-Paczy\'{n}ski (AP) effect\cite{AP}, which is a geometric distortion due to the discrepancy between the true cosmology and the fiducial one used to convert redshifts to distances, is irrelevant at $z=0$\cite{2020JCAP...05..005D}. As studied\cite{Samushia:2021ixs}, the AP effect can make the small-scale bispectrum more informative for constraining the standard ruler than the power spectrum ($P_{\rm pre}$), thus it is worth revisiting the case in which $P_{\rm post}$ and $P_{\rm cross}$ are added to the analysis.

To further demonstrate the efficacy of our method, we perform another analysis at a higher redshift using $P$ and $B_0$ from an independent set of mocks: a suite of $4000$ high-resolution $N$-body mocks ($512^3$ particles in a box with $500 \ h^{-1} {\rm Mpc}$ a side) produced at $z=1.02$. This allows us to include the AP effect when performing the BAO and RSD analysis. This test confirms that $P_{\rm pre}, \ P_{\rm post}$ and $P_{\rm cross}$ are complementary for constraining cosmological parameters, and that $P_{\rm all}$ contains almost all the information in $P$ combined with $B_0$, which is consistent with our findings from the {\sc Molino} analysis (see Methods and Supplement Figures 16-19 for more details).

Stage-IV redshift surveys including the Dark Energy Spectroscopic Instrument (DESI)\cite{DESI}, Euclid\cite{EUCLID} and the Prime Focus Spectrograph (PFS)\cite{PFS} will release galaxy maps over a wide range of { redshifts} with an exquisite precision. As long as the distribution of a tracer in a given redshift range is not too sparse, namely, the number density is not lower than $10^{-4} \ h^{3} {\rm Mpc}^{-3}$ so that a reconstruction can be efficiently performed\cite{White:2010qd}, the method presented in this work can be directly applied to extract high order statistics for constraining cosmological parameters from 2-point measurements, which is computationally much more efficient to perform. Since the reconstruction will be performed anyway for most ongoing and forthcoming galaxy surveys to improve the BAO signal, our proposed analysis can be performed at almost no additional computational cost.

Additional work is required to build a link between cosmological parameters to the full shape of power spectra for a likelihood analysis, and this is challenging using perturbation-theory-based models on (quasi-) nonlinear scales, especially for the reconstructed power spectrum and the cross power spectrum. However, model-free approaches including the simulation-based emulation\cite{2010ApJ...713.1322L, Kobayashi:2020zsw, Neveux:2022tuk}, can be used for performing the $P_{\rm all}$ analysis down to nonlinear scales, in order to extract the cosmological information from the power spectra to the greatest extent. The emulator-based $P_{\rm all}$ analysis was recently performed and validated \cite{PallEmulator:2023}, which well demonstrates the idea proposed in this work.

\newpage

\begin{addendum}

\item[Acknowledgements] We thank Florian Beutler, Shi-Fan Chen, Yipeng Jing, Yosuke Kobayashi, Baojiu Li, Levon Pogosian, Uro\v{s} Seljak, Shun Saito, Atsushi Taruya, Francisco Villaescusa-Navarro, Martin White, Hans Winther, Hanyu Zhang and Pengjie Zhang for discussions. We thank our reviewer Naonori Sugiyama and another three anonymous reviewers for their insightful comments and suggestions, which have significantly improved this work. YW is supported by NSFC Grants (12273048, 11890691), National Key R\&D Program of China (2022YFF0503404, 2023YFA1607800, 2023YFA1607803), the CAS Project for Young Scientists in Basic Research (No. YSBR-092), the Youth Innovation Promotion Association CAS, and the Nebula Talents Program of NAOC. GBZ is supported by the National Key Basic Research and Development Program of China (No. 2018YFA0404503), NSFC Grants 11925303, 11720101004 and 11890691. Research at Perimeter Institute is supported in part by the Government of Canada through the Department of Innovation, Science and Economic Development Canada and by the Province of Ontario through the Ministry of Colleges and Universities. RT is supported by MEXT/JSPS KAKENHI Grant Numbers 20H05855 and 20H04723. Numeric work was performed on the UK Sciama High Performance Computing cluster supported by the ICG, University of Portsmouth.

\item[Author contributions] YW contributed to the idea and the development of the pipeline, performed the analysis on the {\sc Molino} mocks, produced the results and contributed to the draft. GBZ proposed the idea, developed the pipeline, preformed the analysis on the $N$-body mocks and wrote the draft. KK contributed to the idea and pipeline, led the effort on the theoretical interpretation of the result and co-wrote the draft. WJP contributed to the idea, theoretical interpretation and observational implications of this work, and co-wrote the draft. RT and CH provided the $N$-body mocks, performed the reconstruction and measured the power spectra and bispectrum. HGM and CHH contributed to the bispectrum analysis, and to the draft. RZ built a theoretical model for the $k$-dependent noise for the cross-power spectrum, to confirm the measurement using the HSHD approach. WZ, RZ and XM contributed to building an EFT-based theoretical model for the cross-power, which helped with interpretation of the result. YY and HZ provided another set of $N$-body mock for a cross validation, and helped with interpreting the de-correlation effect. FG preformed a test on the eBOSS mocks to confirm the de-correlation effect between $P_{\rm pre}$ and $P_{\rm post}$.

\item[Competing Interests] The authors declare that they have no
competing interest.
\end{addendum}

\clearpage

\begin{figure}
\begin{center}
\includegraphics[width=\linewidth]{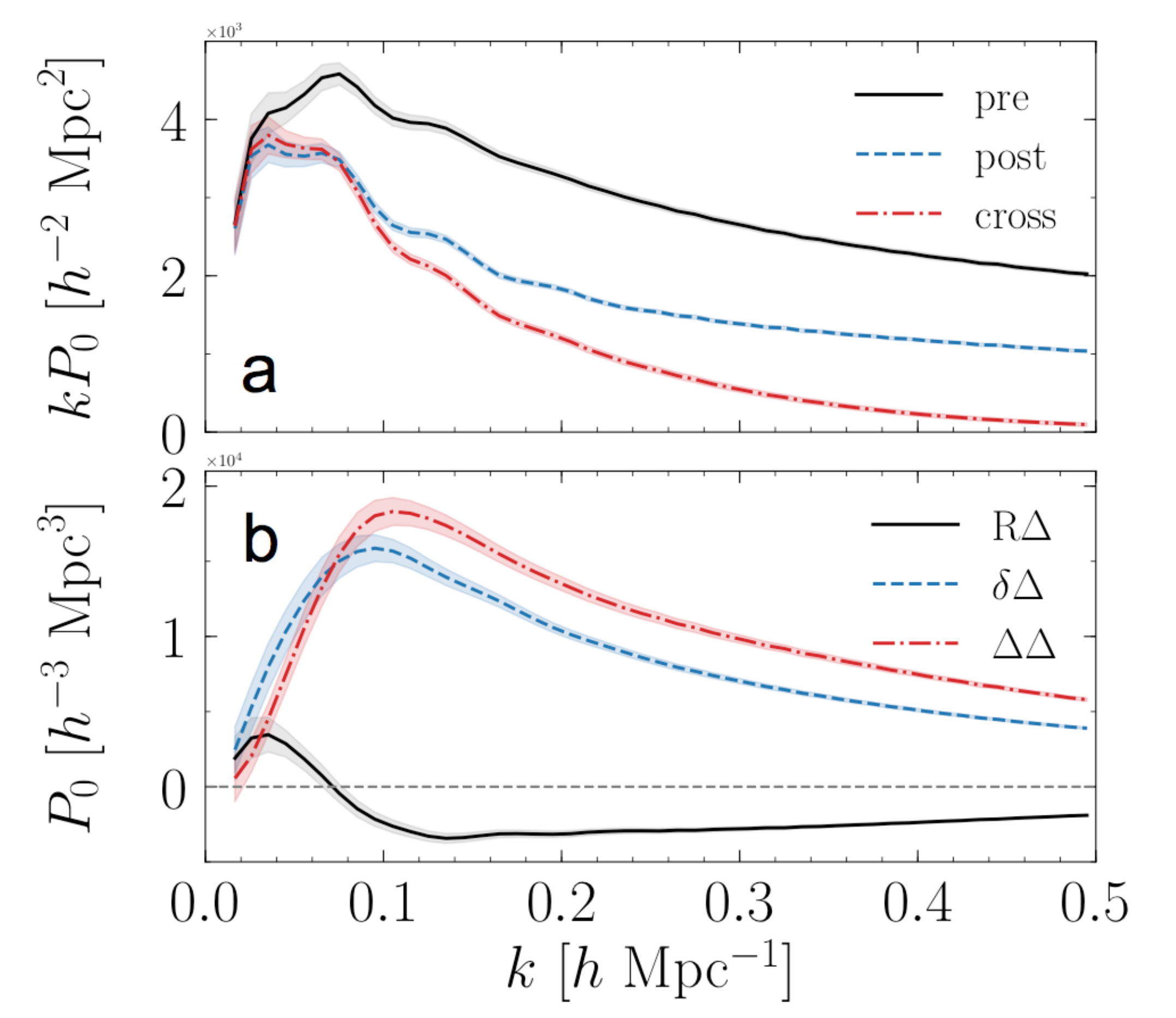}
\label{fig:fig1}
\end{center}
\end{figure} 

\noindent {\bf Figure 1. The measured power spectrum monopole.} \\ Panel {\bf a}: The monopole (multiplied by $k$) of three types of power spectra indicated in the legend, measured from the {\sc Molino} galaxy mocks; {\bf b}: The rotated power spectra monopole defined in Eq.~(\ref{eq:rotated}). In both panels, the lines in the centre denote the mean of the mocks, and the shades represent the 68\% confidence level uncertainty.

\clearpage

\begin{figure}
\begin{center}
\includegraphics[width=\linewidth]{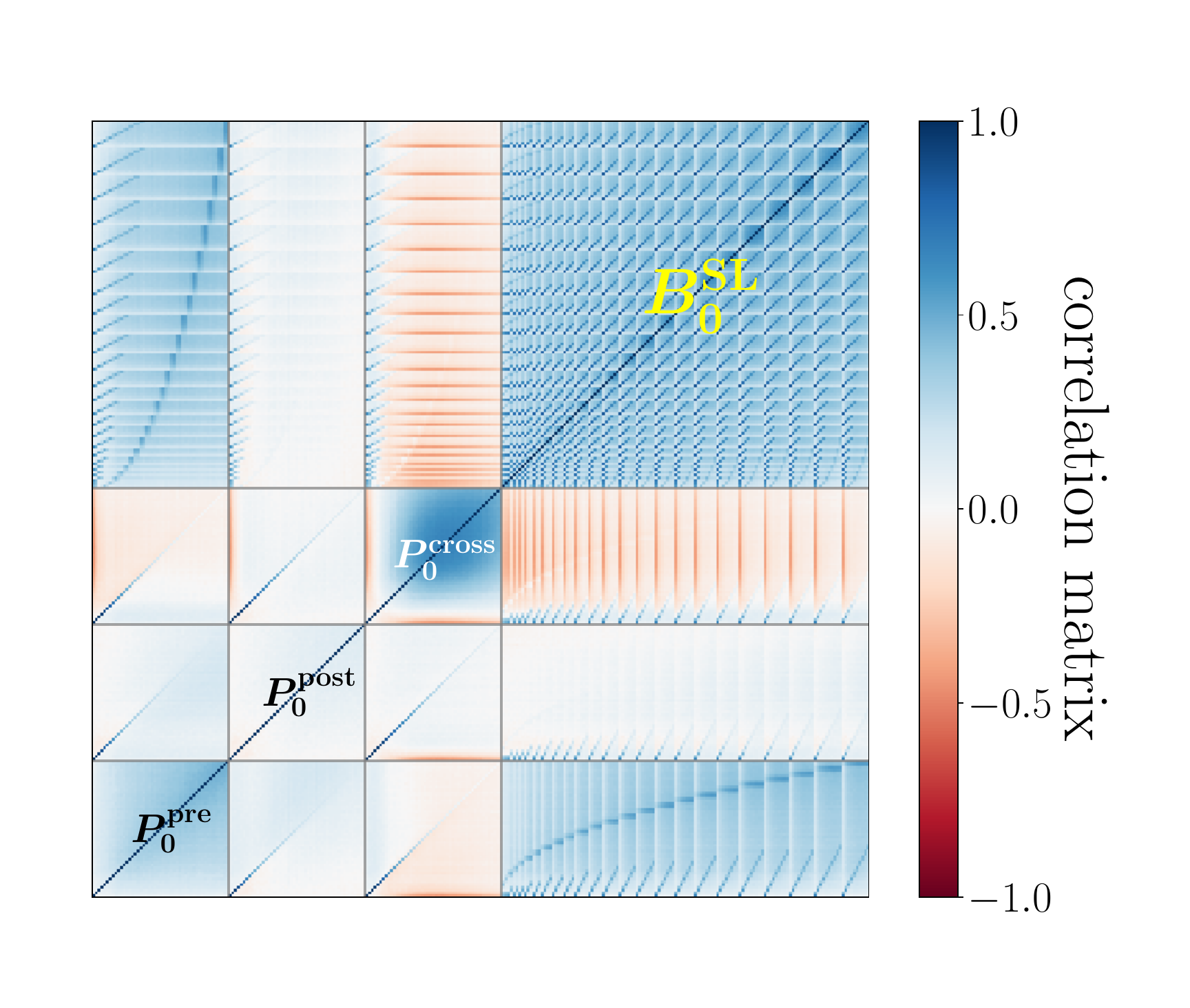}
\end{center}
\end{figure} 

\noindent {\bf Figure 2. Part of the correlation matrix between the power spectra and bispectrum.}\\
The correlation matrix for the monopole of three types of power spectra ($P_0^{\rm pre}, P_0^{\rm post}, P_0^{\rm cross}$), and of the bispectrum in the squeezed limit ($B_0^{\rm SL}$), \ie\,$k_1=k_2 \gg k_3$, derived from the {\sc Molino} galaxy mocks. The horizontal and vertical lines separate each block for visualisation. For all blocks, the associated $k$ or $k_1$ increases from $0.01$ to $0.5~h~{\rm Mpc}^{-1}$, from left to right, and from bottom to top.

\clearpage

\begin{figure}
\begin{center}
\includegraphics[width=\linewidth]{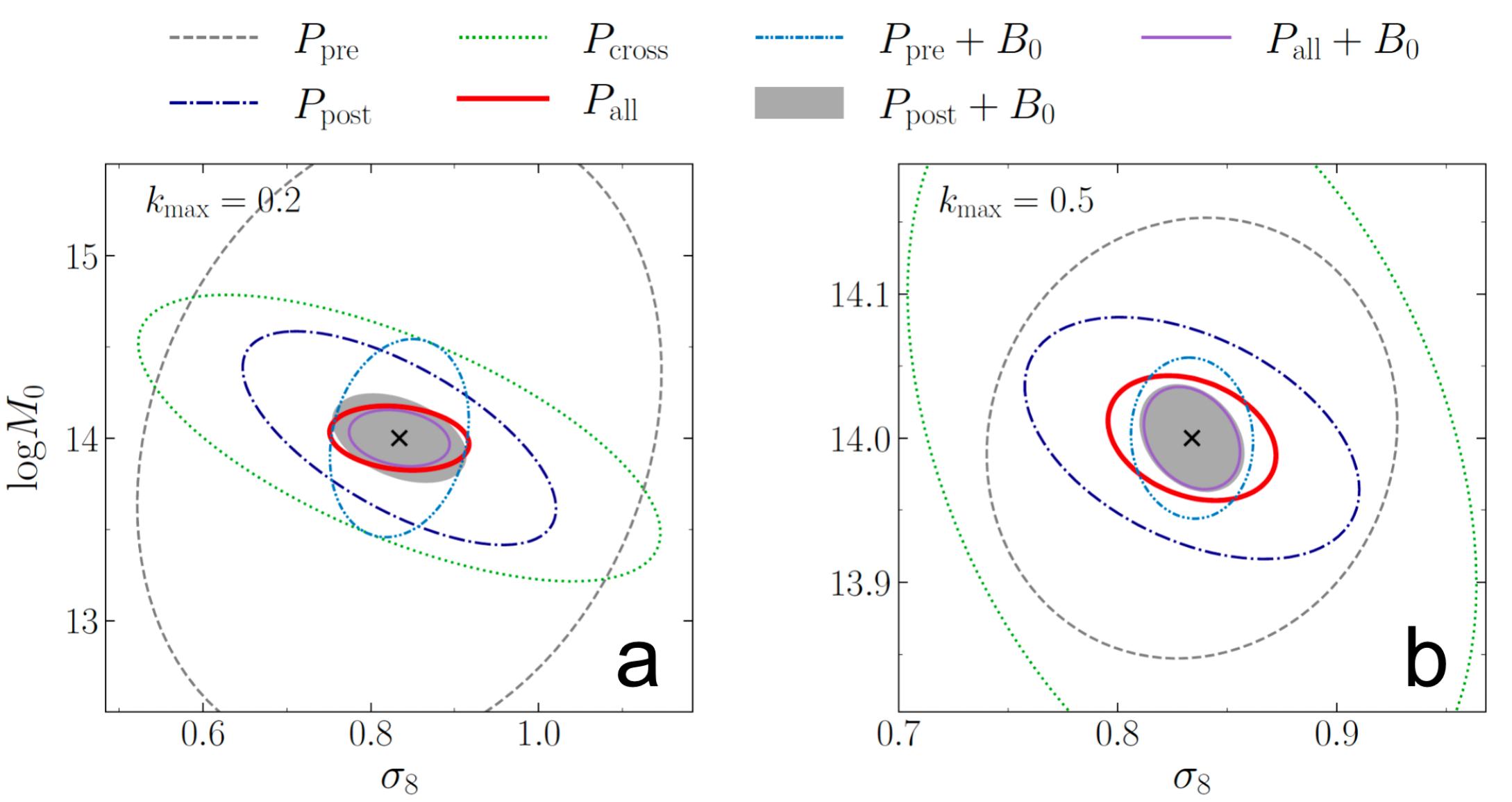}
\end{center}
\end{figure} 

\noindent {\bf Figure 3. The 68\% confidence level contour plots on $\sigma_8$ and ${\rm log}M_0$ derived from various data combinations.} \\
Panels {\bf a} and {\bf b} show the constraint with $k_{\rm max}=0.2$ and $0.5~h~{\rm Mpc}^{-1}$, respectively. In each panel, the constraints are from the pre-reconstructed power spectrum ($P_{\rm pre}$) alone (gray dashed line), post-reconstructed power spectrum ($P_{\rm post}$) alone (dark blue dash-dotted line), cross power spectrum between the pre- and post-reconstructed density fields ($P_{\rm cross}$) alone (green dotted line), the combination of pre-, post-reconstructed and cross-power spectra ($P_{\rm all}$) (red solid line), the combination of pre-reconstructed power spectrum and bispectrum ($P_{\rm pre}+B_{\rm 0}$) (light blue dash-dot-dotted line), the combination of post-reconstructed power spectrum and bispectrum ($P_{\rm post}+B_{\rm 0}$) (gray filled region) and the combination of $P_{\rm all}$ and bispectrum ($P_{\rm all}+B_{\rm 0}$) (purple solid line).

\clearpage

\begin{figure}
\begin{center}
\includegraphics[width=\linewidth]{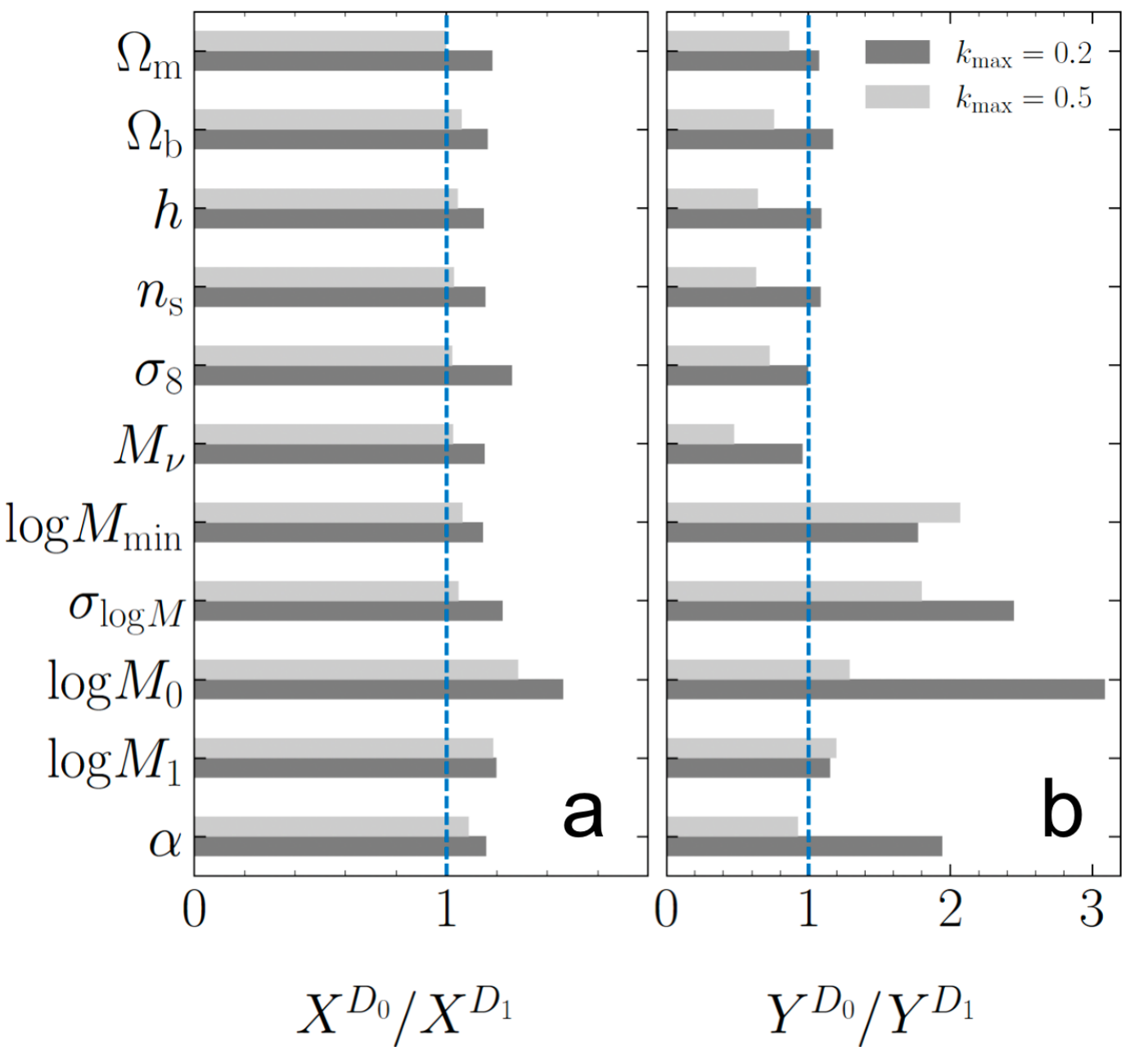}
\end{center}
\end{figure}

\noindent {\bf Figure 4. The Figure of Merit (FoM) of each individual parameter derived from all power spectrum combined, rescaled by those derived from $P_{\rm pre}+B_0$.}\\
The quantities $X$ in panel {\bf a} and $Y$ in panel {\bf b} are defined as the FoM of each individual parameter with or without all other parameters fixed. Specifically, $X\equiv\sqrt{F_{ii}}$ and $Y\equiv1/\sqrt{C_{ii}}$ where $F$ is the Fisher matrix and $C\equiv F^{-1}$. $D_0$ and $D_1$ denote $P_{\rm all}$ and $P_{\rm pre}+B_0$ respectively. The dark and light gray bars in each panel show the cases with $k=0.2$ and $0.5~h~{\rm Mpc}^{-1}$, respectively. The vertical dashed lines show a full recovery of information from dataset $P_{\rm pre}+B_0$. The smoothing scale is set to be $10~h^{-1}~{\rm Mpc}$ when performing the reconstruction.

\clearpage

\begin{table}
\begin{adjustbox}{width=\columnwidth,center}
\begin{tabular}{c|ccccccc|ccccccc}

\hline\hline
                                            & \multicolumn{7}{c|}{$k_{\rm max}=0.2~h~{\rm Mpc}^{-1}$} & \multicolumn{7}{c}{$k_{\rm max}=0.5~h~{\rm Mpc}^{-1}$} \\
 \cline{2-15}                                            
                                            & $P_{\rm pre}$  & $P_{\rm post}$ & $P_{\rm all}$ & \tx{$P_{\rm all}^{\rm SN}$} & $P_{\rm pre} + B_0$ & $P_{\rm post} + B_0$& \tc{$P_{\rm all} + B_0$} & $P_{\rm pre}$  & $P_{\rm post}$ & $P_{\rm all}$ & \tx{$P_{\rm all}^{\rm SN}$}& $P_{\rm pre} + B_0$ & $P_{\rm post} + B_0$ & \tc{$P_{\rm all} + B_0$}\\
\hline
$\sigma({\Omega_{\rm m}})$ & $1$ &$0.66$ &$0.41$ &\tx{$0.44$} &$0.44$ &$0.39$ &$0.32$ & $1$ &$0.86$ &$0.51$ &\tx{$0.52$} &$0.44$ &$0.37$ &$0.33$\\
$\sigma({\Omega_{\rm b}})$ & $1$ &$0.69$ &$0.44$& \tx{$0.47$} &$0.52$ &$0.44$ &$0.37$ & $1$ &$0.76$ &$0.50$& \tx{$0.51$} &$0.38$ &$0.33$ &$0.31$\\
$\sigma(h)$ & $1$ &$0.75$ &$0.43$ & \tx{$0.47$} &$0.47$ &$0.43$ &$0.34$ & $1$ &$0.86$ &$0.48$& \tx{$0.49$}  &$0.31$ &$0.28$ &$0.26$\\
$\sigma({n_{\rm s}})$ & $1$&$0.79$ &$0.41$& \tx{$0.44$} &$0.44$ &$0.42$ &$0.33$ & $1$ &$0.92$ &$0.45$& \tx{$0.45$} &$0.28$ &$0.27$ &$0.25$\\
$\sigma({\sigma_8})$ & $1$ &$0.60$ &$0.27$& \tx{$0.28$} &$0.26$ &$0.25$ &$0.19$ & $1$ &$0.81$ &$0.41$& \tx{$0.41$} &$0.30$ &$0.25$ &$0.23$\\
$\sigma({M_{\nu}~[{\rm eV}]})$ & $1$ &$0.88$&$0.36$&\tx{$0.48$}  &$0.34$ &$0.33$ &$0.25$ & $1$ &$0.84$ &$0.46$& \tx{$0.50$} &$0.22$ &$0.22$ &$0.20$\\\hline
${\rm FoM_{\rm cos}}$ & $1$ &$1.6$ &$2.5$& \tx{$2.6$}  &$2.0$ &$2.4$ &$2.9$ &$1$ &$1.6$ &$2.3$& \tx{$2.4$} &$2.6$ &$3.0$ &$3.2$
\\\hline
$\sigma({{\rm log}M_{\rm min}})$ & $1$&$0.66$ &$0.21$& \tx{$0.10$} &$0.38$&$0.39$ &$0.14$ & $1$ &$0.54$ &$0.19$& \tx{$0.14$} &$0.39$ &$0.31$ &$0.13$\\
$\sigma({\sigma_{{\rm log}_{M}}})$ & $1$ &$0.59$&$0.18$ & \tx{$0.16$} &$0.43$ &$0.38$ &$0.14$ & $1$ &$0.50$&$0.16$& \tx{$0.17$} &$0.30$ &$0.21$ &$0.12$\\
$\sigma({{\rm log}M_{0}})$ & $1$ &$0.33$&$0.10$&\tx{$0.10$}  &$0.31$ &$0.14$ &$0.09$ & $1$ &$0.55$&$0.28$ &\tx{$0.28$} &$0.37$ &$0.24$ &$0.23$\\
$\sigma({{\rm log}M_{1}})$ & $1$ &$0.89$&$0.27$& \tx{$0.15$} &$0.31$ &$0.36$ &$0.22$ & $1$ &$0.96$&$0.43$& \tx{$0.29$} &$0.51$ &$0.48$ &$0.35$\\
$\sigma({\alpha})$ & $1$ & $0.54$ &$0.14$&\tx{$0.12$} &$0.27$ &$0.20$ &$0.12$ & $1$ &$0.94$ &$0.41$& \tx{$0.28$} &$0.38$ &$0.34$ &$0.32$\\\hline
${\rm FoM_{\mathbf{\Theta}}}$ & $1$ &$1.6$ &$2.7$ & \tx{$3.4$} &$1.9$ &$2.4$ &$3.2$ &$1$ &$1.4$ &$2.1$ & \tx{$2.4$} &$2.0$ &$2.3$ &$2.8$ \\
\hline\hline

\end{tabular}
\label{table}

\end{adjustbox}

\caption{{\bf A quantification of the information content extracted from various observables measured from the {\sc Molino} mocks.} 
The 68\% confidence level uncertainty of parameters derived from various datasets, with two different choices of the maximal wavenumber $k_{\rm max}$. The column $P_{\rm all}^{\rm SN}$ shows the result derived from $P_{\rm all}$ with the shot-noise component of $P_{\rm cross}$ kept. The last row, ${\rm FoM}_{\Theta}$ shows the Figure-of-Merit (FoM), defined as $\left[{\bf det} (F)\right]^{1/(2N_{\rm p})}$, where $F$ denotes the Fisher matrix and $N_{\rm p}$ is the total number of free parameters. The row of ${\rm FoM}_{\rm cos}$ shows the same FoM as defined above, but with all HOD parameters fixed. All quantities are normalised by that of the $P_{\rm pre}$ result.}
\end{table}%

\clearpage


\section{Methods}

\subsection{The mock catalogs $-$ {\sc Molino} galaxy mocks at $z=0$ }

The {\sc Molino} catalogs\cite{Molino} are a suite of publicly available galaxy mock catalogs that 
were constructed  to quantify the total cosmological information content of different galaxy 
clustering observables using Fisher matrix forecasting. They are constructed from the {\sc Quijote} suite of $N$-body simulations\cite{Quijote} using the halo occupation distribution (HOD) framework. HOD provides a statistical prescription for populating dark matter halos with central and satellite galaxies and has been successful in reproducing a wide range of observed galaxy clustering statistics. In particular, the {\sc Molino} catalogs use the standard HOD model\cite{Zheng07}, which has five free parameters: $\left\{\right.\log M_{\rm min},\allowbreak \sigma_{\log M},\allowbreak \log M_0,\allowbreak \alpha,\allowbreak \log M_1 \left.\right\}$. {\sc Molino} includes $15,000$ galaxy catalogs that are constructed at a fiducial set of cosmological parameters ($\Omega_m = 0.3175, \Omega_b = 0.049, h = 0.6711, n_s = 0.9624, \sigma_8=0.834$, $M_\nu = 0$) and HOD parameters ($\log M_{\rm min} = 13.65, \sigma_{\log M} = 0.2, \log M_0 = 14.0, \alpha = 1.1, \log M_1 = 14.0$), which are based on the best-fit HOD parameters for the SDSS \tc{$M_r<-21.5$} and $-22$ samples\cite{Zheng07}. 


The $15,000$ {\sc Molino} mocks for the fiducial cosmology are designed for accurately estimating the covariance matrices of the galaxy clustering observables, including the power spectra and bispectra. In addition, a separate set of the {\sc Molino} mocks are produced for estimating the derivatives with respect to cosmological parameters (including the HOD ones) using the finite difference method (see Supplement Note 2 for details). For this purpose, $60,000$ galaxy mocks are constructed at $24$ cosmologies that are slightly different from the fiducial one\cite{Molino}.

Since the data covariance matrices and the derivatives are all evaluated numerically using mocks, it is important to ensure that the result derived from the Fisher matrix approach is robust against numerical issues, as argued in\cite{Hahn:2019zob,Molino,Coulton:2022rir,Paillas:2022wob}. We therefore perform numerical tests to check the dependence of our Fisher matrix calculation on the number of mocks, and find that the marginalised uncertainties of all the concerning parameters are well converged given the number of mocks available. The details are presented in the Supplement Note 2.

\subsection{The mock catalogs $-$ $4000$ high-resolution $N$-body mocks at $z=1.02$} 

To confirm our findings from the {\sc Molino} mocks, we perform an independent mock test on a suite of $4000$ high-resolution $N$-body simulations with $512^3$ dark matter particles in a $L=500~h^{-1}{\rm Mpc}$ box at $z=1.02$\cite{Hikage2020}. The fiducial cosmology used for this set of mocks is consistent with the Planck 2015\cite{Planck2015} observations.

\subsection{The mock catalogs $-$ {\sc COLA} mocks at multiple redshifts}

To investigate how the decorrelation between $P_{\rm pre}$ and $P_{\rm post}$ varies with redshifts, we perform another set of N-body simulations using the COmoving Lagrangian Acceleration ({\sc COLA})\cite{Tassev:2013pn} method with the {\tt MG-PICOLA} code\cite{Winther:2017jof}. The mocks are performed using $256^3$ dark matter particles in a $L=256~h^{-1}{\rm Mpc}$ box, and snapshots at $z=0,1,2,3,5,10,15$ are analysed, to cover a sufficiently wide range of redshifts. Although the COLA mocks are approximate, the accuracy and reliability has been well demonstrated in the literature\cite{Tassev:2013pn,Winther:2017jof,Howlett:2015hfa}.

\subsection{The reconstruction process}

An anisotropic reconstruction\cite{Chen19} is performed on each realisation of the {\sc Molino} galaxy mocks with two choices of the smoothing scale, $S=10$ and $20~h^{-1}~{\rm Mpc}$ (All results presented in the main text are for the $10~h^{-1}~{\rm Mpc}$ case, while results for $20~h^{-1}~{\rm Mpc}$ are shown in the Supplement information). Specifically, a smoothing is performed by convolving density field with the kernel $K(k)={\rm exp}\left[- (k\ S)^2/2\right]$ in Fourier space. Note that in this procedure the information on scales below the smoothing scales gets erased, and there are studies on choosing the proper smoothing scale\cite{Seo15}. In principle, the smoothing scale can be made sufficiently small to restore more information, for example, no smoothing is needed at all in the nonlinear reconstruction methods\cite{NLrecon}, and we will apply our pipeline to those reconstruction schemes for further investigation. After the smoothing, the displacement vector is solved using the Zeldovich approximation, \ie, $\tilde{\mathbf{s}}(\mathbf{k})=-\frac{i\mathbf{k}}{k^2}\frac{\delta({\mathbf{k}})}{b_{\rm in}+f_{\rm in} \mu^2}K(k)$, where $\delta$ denotes the nonlinear redshift-space overdensity, $b_{\rm in}$ and $f_{\rm in}$ are the {\it input} linear bias and the logarithmic growth rate for the density field, respectively. Note that $\{b_{\rm in}, f_{\rm in}\}$ does not have to be identical to the true underlying $\{b, f\}$ of the density field, thus they are {\it not} free parameters to be determined. The post-reconstructed power spectrum for a given $\{b_{\rm in}, f_{\rm in}\}$ can be modeled using either the perturbation theory\cite{wrongrecon}, or an emulation approach, as developed in Wang \etal\cite{PallEmulator:2023} It is true that an inappropriate choice of $\{b_{\rm in}, f_{\rm in}\}$, \eg, a set of $\{b_{\rm in}, f_{\rm in}\}$ that is significantly different from the truth, may affect the efficiency of the BAO reconstruction, but the impact from using $\{b_{\rm in}, f_{\rm in}\}$ can be well modeled and corrected for, so this process is not expected to generate bias or uncertainties.

To demonstrate that the result would not get biased by an inappropriate set of $\{b_{\rm in}, f_{\rm in}\}$, Wang \etal\cite{PallEmulator:2023} uses a significantly wrong set of $\{b_{\rm in}, f_{\rm in}\}$ for the reconstruction, namely, $\{b_{\rm in}=0.9 b, f_{\rm in}=0.7 f\}$, where $\{b, f\}$ are the true $b$ and $f$ of the density field. This level of deviation from the true value is greater than $3\sigma$ level, given the uncertainty of $b$ and $f$ constrained by the BOSS (DR12) survey\cite{BOSSDR12}. The impact of using such a wrong set of $\{b_{\rm in}, f_{\rm in}\}$ is corrected for by the properly trained emulator, and as demonstrated in Fig. 6 of Wang \etal\cite{PallEmulator:2023}, using this set of $\{b_{\rm in}, f_{\rm in}\}$ does not bias, or dilute the final parameter constraint. In summary, it is expected that the choice of $\{b_{\rm in}, f_{\rm in}\}$ used in this work does not bias the result, and a more in-depth assessment on the potential influence of the choice on $\{b_{\rm in}, f_{\rm in}\}$ is left for a future study on a joint $P_{\rm all}$ analysis using the actual observational data.

 An inverse Fourier transformation on $\tilde{\mathbf{s}}$ returns the configuration-space displacement field $\mathbf{s}(\mathbf{x})$, which is used to move both the galaxies and randoms. We also perform the anisotropic Lagrangian reconstruction\cite{Seo15} on each realisation of the $N$-body mocks, but only with a smoothing scale $S=10~h^{-1}{\rm Mpc}$.

Note that the information content in the reconstructed power spectrum is the same no matter whether the RSD is kept or not during the reconstruction process, and we have numerically confirmed this by performing the analysis with the isotropic reconstruction\cite{Seo15}, in which the RSD is removed using the fidicual $f$ and $b$ used for producing the mocks.

Also note that the BAO reconstruction procedure is not always required for extracting geometric information in the galaxy clustering. For example, when using the information in the linear point \cite{LP1, LP2, LP3, LP4}, no reconstruction is required. Also, the estimated $\alpha$ from the traditional BAO methods and from the linear point approach may conceptually differ, and a comparison is beyond the scope of this work.

\subsection{Measurement of the power spectrum multipoles}

The multipoles (up to $\ell=4$) of both the pre- and post-reconstructed density fields are measured using an FFT-based estimator\cite{Hand17} implemented in {\tt N-body kit}\cite{nbodykit}. The shot-noise, which reflects the discreteness of the density field, is removed as a constant for the monopole of the auto-power. The $k$-binning is $\Delta k = 0.01~h{\rm Mpc}^{-1}$ for both the {\sc Molino} and $N$-body mocks.

Care needs to be taken when measuring the cross-power spectrum between the pre- and post-reconstructed density fields, since the raw measurement using the FFT-based estimator is contaminated by a scale-dependent shot-noise: on large scales, the post-reconstructed field resembles the unreconstructed one, making the cross-power spectrum essentially an auto-power, thus it is subject to a shot-noise component. On small scales, however, the shot-noise largely drops because the two fields effectively decorrelate. 

To obtain a measured cross-power spectrum whose mean value reflects the true power spectrum in the data such that no subtraction of the noise component is required, we adopt the half-sum and half-difference (HS-HD) approach\cite{noise}. We start by randomly dividing the catalog into two halves, dubbed $\delta_1$ and $\delta_2$, and the corresponding reconstructed density fields are ${\rm R_1}$ and ${\rm R_2}$, respectively.

Let \ba
{\rm HS}\equiv\frac{\delta_1+\delta_2}{2},~~{\rm HD}\equiv\frac{\delta_1-\delta_2}{2},
\ea and 
\ba
{\rm HS^R}\equiv\frac{\rm R_1+\rm R_2}{2},~~{\rm HD^R}\equiv\frac{\rm R_1-\rm R_2}{2}.
\ea
Then $\rm HS^{(R)}$ contains both the signal and noise, but $\rm HD^{(R)}$ only contains the noise. Hence the \tc{cross-power spectrum estimator is},
\be
 \hat{P}_{\rm cross} = \langle \rm HS, HS^R \rangle-\langle \rm HD, HD^R \rangle
 = \frac{\langle \delta_1, R_2 \rangle+\langle  \delta_2, R_1 \rangle}{2}.
\ee

The scatter of $\hat{P}_{\rm cross}$ around the mean value allows for an estimation of the covariance matrix, which is a 4-point function\cite{Sugiyama:2019ike},  shown in Figure 2. By comparing $\hat{P}_{\rm cross}$ with that measured without splitting the samples, we can obtain the noise power spectrum, as shown in Supplement Figures 14 and 15 (for cases with $S=10$ and $20~h^{-1}~{\rm Mpc}$, respectively), which is apparently scale-dependent. The noise is anisotropic, and thus it affects even for multipoles with $\ell\ne0$.

Since a change in HOD parameters can result in a change in the number density of the galaxy sample and thus affect the shot-noise, the shot-noise can in principle be used to constrain the HOD parameters. We therefore perform an additional Fisher projection with the shot-noise kept in the spectra, and find that the constraints on HOD parameters can be improved in general, but the constraint on cosmological parameters is largely unchanged (see the $P_{\rm all}^{\rm SN}$ column in Table 1).

\subsection{Measurement of the bispectrum monopole}

We measure the galaxy bispectrum monopole, $B_0$, for all of the mock catalogs using the publicly available $\mathtt{pySpectrum}$ package\cite{Hahn:2019zob,Molino}. Galaxy positions are first interpolated onto a grid using a fourth-order interpolation scheme and then Fourier transformed to obtain $\delta(k)$. 
Afterwards $B_0$ is estimated using 
\begin{equation}  
B_0(k_1, k_2, k_3) = \frac{1}{V_B} \int\limits_{k_1}{\rm d}^3q_1
\int\limits_{k_2}{\rm d}^3q_2 \int\limits_{k_3}{\rm d}^3q_3~\delta_{\rm
D}({\bf q_{123}})~\delta({\bf q_1})~\delta({\bf q_2})~\delta({\bf q_3}) -
B^{\rm SN}_0
\end{equation}
where $\delta_D$ is the Dirac delta function, $V_B$ is the normalization factor
proportional to the number of triplets that can be found in the $k_1, k_2, k_3$
triangle bin, and $B^{\rm SN}_0$ is the Poisson shot noise correction term.
Triangle configurations are defined by $k_1, k_2, k_3$, and for the {\sc Molino} mocks, the width of the bins is
$\Delta k = 3 k_f$, where $k_f = 2\pi/(1000~h^{-1}{\rm Mpc})$, and for the $N$-body mocks, $\Delta k = 0.02~h{\rm Mpc}^{-1}$.

\subsection{An AP test performed on the {\sc Molino} mocks}

Although the AP effect plays no role for the {\sc Molino} mock since it is produced at $z=0$\cite{2020JCAP...05..005D}, we perform a test by isotropically stretching the scales and angles using pairs of AP parameters calculated at a non-zero redshift. This gives us an idea about whether this artificial and exaggerated AP effect can change the main conclusion of this work that the cosmological information content in $P_{\rm all}$ is almost the same as or more than that in $P_{\rm pre}+B_0$. In practice, we use {the} ($\alpha_{||}$ and $\alpha_{\bot}$) pairs computed at $z_{\rm eff}=0.5$ and $1.0$ respectively to stretch the wave numbers along and across the line of sight directions, and repeat the analysis. As shown in the Supplement Table (see Supplement Note 3), this added `artificial' AP effects can generally tighten the constraint, but the relative constraints from $P_{\rm all}$ and $P_{\rm pre}+B_0$ are largely unchanged, meaning that the main conclusion of this paper remains the same if the AP effect is taken into account.

\subsection{An AP test on the $N$-body mocks}

We perform an additional Fisher matrix analysis\cite{Fisher} on the AP parameters using $4000$ realisations of $N$-body particle mocks produced at $z=1.02$ in redshift space. Part of the observables (the power spectrum monopole) are shown in Supplement Figure 16. From panel {\bf a} we see that the amplitude of $P_{\rm cross}$ decreases dramatically with scales, indicating a decorrelation between $P_{\rm pre}$ and $P_{\rm post}$ below quasi-nonlinear scales, which is confirmed by the correlation coefficient (the normalised covariance) plotted in panel {\bf b}. This decorrelation, which is not caused by the shot noise given the negligible noise level in the mocks, is a clear evidence of the complementarity among the power spectra.

The cumulative signal-to-noise ratio (SNR) is shown in panel {\bf a} of Supplement Figure 17, in which we see that $P_{\rm all}$ is more informative than $P_{\rm pre}$, and that $P+B_0$ has slightly higher SNR on small scales.

We first perform an AP test on the isotropic dilation parameter ${ \alpha_{\rm iso}}$, which is defined as the ratio of the true spherically-averaged scale of the standard ruler to the fiducial one. This dilation parameter depends on cosmological parameters, and can be constrained using the monopole of the power spectrum and bispectrum. The wavenumber $k$ gets dilated by ${ \alpha_{\rm iso}}$ due to the AP effect, thus the observables are,
\ba
P_0^{T}(k')&=&A_0\frac{1}{\alpha_{\rm iso}^3}P_0^{T} \left(k / \alpha_{\rm iso} \right)\, \\
B_0(k_1',k_2',k_3')&=&A_B\frac{1}{\alpha_{\rm iso}^6}B_0 (k_1 / \alpha_{\rm iso}, k_2 / \alpha_{\rm iso},k_3 / \alpha_{\rm iso}),
\ea 
where $T$ denotes the type of $P_0$, namely, $T=\{{\rm pre}, {\rm post}, {\rm cross}\}$, and the parameters $A_0$ and $A_B$ are used to parameterize the overall amplitudes of power spectrum monopole and bispectrum monopole, respectively. Since the purpose of this test is to study the impact of AP parameters, the relevant parameters are \{$\alpha_{\rm iso}$, $\ln A_0$, $\ln A_B$\}, and these are free parameters in this calculation. Other parameters are held fixed to avoid confusion. The derivative with respect to ${\alpha_{\rm iso}}$ is evaluated semi-analytically as
\ba 
\frac{\partial P_0^{T}}{\partial \alpha}  &=& -3P_0^{T} -  \frac{d P_0^{T}}{d \ln k},\\
\frac{\partial B_0}{\partial \alpha_{\rm iso}}  &=& -6B_0 - \left( \frac{\partial B_0}{\partial \ln k_1} +\frac{\partial B_0}{\partial \ln k_2}  +\frac{\partial B_0}{\partial \ln k_3} \right).
\ea
Then the constraint on $\alpha_{\rm iso}$ is derived after marginalising over the amplitudes $A_0$ and $A_B$, and it is shown in panel {\bf b} of Supplement Figure 17. The FoM of $\alpha_{\rm iso}$ shows up step-like features due to the BAO feature, as previously discovered\cite{Samushia:2021ixs}, and $P_{\rm all}$ offers the greatest FoM, until overtaken by $P+B_0$ at $k_{\rm max}\gtrsim0.37 \ h {\rm Mpc}^{-1}$. 

\tc{We use the first three even multipole moments to assemble the two-dimensional power spectrum, \ie,
\ba
P(k, \mu) \simeq \sum_{\ell=0,2,4} P_{\ell} (k) \mathcal{L}_{\ell}(\mu),
\ea
The bispectrum is similarly assembled using the first three even multipoles with $m=0$\cite{Scoccimarro:1999ed}, which are the most informative ones\cite{Gagrani:2016rfy}, \ie, 
\ba
B\left(k_{1}, k_{2}, k_{3}, \mu_1, \mu_2 \right) \simeq \sum_{\ell=0,2,4} B_{\ell, m=0}\left(k_{1}, k_{2}, k_{3}\right) Y_{\ell,m=0}(\theta, \phi).
\ea 
The wave-number $k_i$ and the cosine of the angle to the line-of-sight $\mu_i$ are stretched by two dilation parameters $\alpha_{\perp}$ and $\alpha_{||}$ due to the AP effect\cite{Ballinger:1996cd, Gil-Marin:2016wya},
\ba q_{i} =\frac{k_{i}}{\alpha_{\perp}}\left[1+\mu_{i}^{2}\left(\frac{1}{F^{2}}-1\right)\right]^{1 / 2},~~\nu_{i} =\frac{\mu_i}{F}\left[1+\mu_i^{2}\left(\frac{1}{F^{2}}-1\right)\right]^{-1 / 2},~~ F = \alpha_{||}/\alpha_{\perp}. \ea
The power spectrum multipoles (the index for the type is omitted for brevity) and bispectrum monopole including the AP effect are respectively given as, 
\ba P_{\ell}(k) &=&\frac{(2 \ell+1)}{2 \alpha_{\perp}^{2} \alpha_{\|}} \int_{-1}^{1} d \mu P\left(q, \nu \right) \mathcal{L}_{\ell}(\mu),\\
B_0(k_1,k_2,k_3) &=& \frac{1}{4\pi \alpha_{\perp}^{4} \alpha_{\|}^2} \int_{-1}^{1} d \mu_1  \int_{0}^{2 \pi}  d \phi B\left(q_{1}, q_{2}, q_{3}, \nu_1, \nu_2 \right).
\ea The free parameters are $\{\alpha_{\perp}, \alpha_{||}, \ln A_{\ell}, \ln A_B\}$, where $A_{\ell} (\ell=0,2,4)$ denotes the overall amplitudes of the power spectrum multipoles, and $A_B$ is the amplitude of the bispectrum monopole. The derivatives with respect to the parameters $\alpha_{\perp}$ and $\alpha_{||}$ are evaluated numerically by 
\ba
\frac{\partial \boldsymbol{O}}{\partial \alpha_{i}} = \frac{\boldsymbol{O}(\alpha_{i}^+) -\boldsymbol{O}(\alpha_{i}^-)}{2 \Delta \alpha_{i}},
\ea where $\boldsymbol{O} \in \{P_{\ell},B_0\}$ denotes the observables, and the step size $\Delta \alpha_{i}=0.01$. Then the constraints on $\alpha_{\perp}$ and $\alpha_{||}$ are derived after marginalising over the amplitudes $A_{\ell}$ and $A_B$.}

\tc{The FoM for $\alpha_{\perp}, \alpha_{||}$ is shown in panel {\bf c} of Supplement Figure 17, and it shows a similar trend as FoM($\alpha_{\rm iso}$). The contour plot for $\alpha_{\perp}, \alpha_{||}$ with $k_{\rm max}=0.4 \ h {\rm Mpc}^{-1}$ is shown in Supplement Figure 18, further highlighting the strong constraining power of $P_{\rm all}$ in comparison to that of $P+B_0$.}

\subsection{A joint BAO and RSD analysis on the $N$-body mocks}

In addition to $\alpha_{\perp}, \alpha_{||}$, we add one more parameter to the analysis, which is $\Delta v$, the parameter describing the change of velocities along the line of sight. This parameter mimics the change of the linear growth rate on large scales, but it also changes the velocity of particles coherently on small scales. We compute the derivatives with respect to $\Delta v$ numerically.

The projection onto the parameters, shown in Supplement Figure 19, demonstrates the advantage of performing a joint analysis using $P_{\rm pre}$, $P_{\rm post}$ and $P_{\rm cross}$. On large scales, $P_{\rm pre}$ and $P_{\rm post}$ are both determined by the linear density field, making the power spectra highly correlated. As shown in panels ${\textbf c_1}$ and ${\textbf c_5}$, the contours derived from $P_{\rm pre}$ and $P_{\rm post}$ have similar orientations and we do not gain by combining them. For $k>0.15~h~{\rm Mpc}^{-1}$, the correlation between  $P_{\rm pre}$ and $P_{\rm post}$ decreases as the pre-reconstructed density field is dominated by the non-linear field while $P_{\rm post}$ still retains the correlation with the linear density field. The contours shown in lines in panel ${\textbf c_7}$, which are derived from power spectra in the $k$ range of $[0.2,0.25]~h~{\rm Mpc}^{-1}$, are almost orthogonal to each other, making the constraint from the combined spectra, as illustrated in the shaded region, significantly tightened. On smaller scales, the post-reconstructed density field is also dominated by the non-linear field and the orientations of the contours are again aligned and the complementarity on smaller scales weakens. This shows that the level of nonlinearity in the power spectrum determines the degeneracies between parameters. Since $P_{\rm pre}$ and $P_{\rm post}$ are affected by different levels of nonlinearities on a given scale, which gives rise to different  degeneracies, a joint analysis using both $P_{\rm pre}$ and $P_{\rm post}$ (and $P_{\rm cross}$) can yield a better constraint by breaking the degeneracies.


\begin{addendum}
 \item[Data Availability] The data that support the plots within this paper and other findings of this study are available from the corresponding author upon reasonable request.
\end{addendum}

\clearpage

{\bf \large Supplementary Information for ``Extracting high-order cosmological information in galaxy surveys with power spectra"}

\author{Yuting Wang$^{1,2}$,
Gong-Bo Zhao$^{1,2,3}$,
Kazuya Koyama$^{4}$,
Will J. Percival$^{5,6,7}$,
Ryuichi Takahashi$^{8}$,
Chiaki Hikage$^{9}$,
H\'ector Gil-Mar\'{\i}n$^{10}$,
ChangHoon Hahn$^{11}$,
Ruiyang Zhao$^{1,3}$,
Weibing Zhang$^{1,3}$,
Xiaoyong Mu$^{1,3}$,
Yu Yu$^{12}$,
Hong-Ming Zhu$^{13}$,
Fei Ge$^{14}$ \\
$^{1}$ National Astronomy Observatories, Chinese Academy of Sciences, Beijing, 100101, P.R.China \\
$^{2}$ Institute for Frontiers in Astronomy and Astrophysics, Beijing Normal University, Beijing 102206, China\\
$^{3}$ School of Astronomy and Space Science, University of Chinese Academy of Sciences, Beijing 100049, P.R.China\\
$^{4}$ Institute of Cosmology and Gravitation, University of Portsmouth, Dennis Sciama Building, Portsmouth PO1 3FX, United Kingdom\\
$^{5}$ Waterloo Centre for Astrophysics, University of Waterloo, 200 University Ave W, Waterloo, ON, N2L 3G1, Canada\\
$^{6}$ Department of Physics and Astronomy, University of Waterloo, 200 University Ave W, Waterloo, ON, N2L 3G1, Canada\\
$^{7}$ Perimeter Institute for Theoretical Physics, 31 Caroline St. North, Waterloo, ON, N2L 2Y5, Canada\\
$^{8}$ Faculty of Science and Technology, Hirosaki University, 3 Bunkyo-cho, Hirosaki, Aomori 036-8561, Japan\\
$^{9}$ Kavli Institute for the Physics and Mathematics of the Universe (Kavli IPMU, WPI), University of Tokyo, 5-1-5 Kashiwanoha, Kashiwa, Chiba, 277-8583, Japan\\
$^{10}$ ICC, University of Barcelona, IEEC-UB, Mart\'{\i} i Franqu\`es, 1, E-08028 Barcelona, Spain\\
$^{11}$ Department of Astrophysical Sciences, Princeton University, Peyton Hall, Princeton NJ 08544, USA\\
$^{12}$ Department of Astronomy, Shanghai Jiao Tong University, Shanghai, 200240, P.R.China\\
$^{13}$ Canadian Institute for Theoretical Astrophysics, University of Toronto, 60 St. George Street, Toronto, Ontario M5S 3H8, Canada\\
$^{14}$ Department of Physics and Astronomy, University of California, Davis, California 95616, USA
}

\newpage
\noindent {\bf Supplementary Note 1. The information content in $P_{R\Delta}$ and $P_{\Delta\Delta}$}

At the leading order in perturbation theory, assuming that the reconstructed field is a linear field, then $R$ and $\Delta$ can be expanded in terms of the linear overdensity $\delta_{\rm L}$ as, $R = \delta_{\rm L}; \ \Delta = F_2 \delta_{\rm L}^2 + F_3 \delta_{\rm L}^3$, where $F_2$ and $F_3$ are the second- and third-order kernel\cite{2017PhRvD..96d3513H}. $P_{R\Delta}$ and $P_{\Delta \Delta}$ are given by 
\begin{eqnarray}
P_{R\Delta}(k) & \propto & P_{\rm L}(k) \int {\bf dp} \ F_3({\bf k}, {\bf p}, {\bf -p}) P_{\rm L}(p) \\
P_{\Delta \Delta}(k) & \propto &  \int {\bf dp} \left[F_2({\bf p}, {\bf k-p})\right]^2 P_{\rm L}(p) P_{\rm L}(|{\bf k-p}|)
\end{eqnarray} where $P_{\rm L}(k)$ is the linear power spectrum. Since the leading-order bispectrum and trispectrum are determined by $F_2$ and $F_3$ respectively, $P_{\Delta \Delta}$ contains the information of the bispectrum at the leading order.

\clearpage

\noindent {\bf Supplementary Note 2. The convergence of the Fisher matrix}

The Fisher matrix is formed as follows, \be F = D^T C^{-1} D, \nonumber \ee
where $D$ and $C$ are the data vector storing the numerical derivatives, and the data covariance matrix, respectively, thus the convergence of $F$ relies on the convergence of $D$ and $C^{-1}$ against the number of mocks. We therefore check the convergence of the marginalised uncertainties of cosmological parameters, as $D$ and $C^{-1}$ are estimated using different number of mocks.

Panel $\bf a$ in Supplement Figure 5 demonstrates the excellent convergence of $F$ against $C^{-1}$ (with $D$ estimated using all the mocks), with the correction factors applied to remove the bias. We then check the convergence of $F$ against $D$ (with $C^{-1}$ estimated using all the mocks). We pay particular attention to the Fisher element for the neutrino mass, because it is recommended to estimate using four pieces of the power spectrum (or bispectrum), shown in Eq. ({\ref{eq:Dmnudef}})\cite{Quijote}, instead of two for other parameters, thus $F_{M_\nu M_\nu}$ may be the most affected Fisher element by numerical issues, if there are any.
\be \label{eq:Dmnudef} \frac{\partial \vec{S}}{\partial M_{\nu}} \sim \frac{\vec{S}(4M_{\nu}) - 12\vec{S}(2M_{\nu}) + 32 \vec{S}(2M_{\nu})- 21 \vec{S}(M_{\nu}=0) }{12 M_{\nu} },\ee where $M_{\nu}$ should be sufficiently small so that the finite difference approaches the derivative. In practice $M_{\nu}$ is set to $0.1$ eV for the Molino mocks. Note that the way the terms are assembled in Eq. ({\ref{eq:Dmnudef}}) is to remove the second order terms in the Taylor series. However, it is not the only arrangement that is free from the second order terms so may not be optimal for removing the numerical noises. The most general form for $\frac{\partial \vec{S}}{\partial M_{\nu}}$ that is second-order terms free is,
\be  \label{eq:Dmnunew} \frac{\partial \vec{S}}{\partial M_{\nu}} \sim \frac{X+\lambda Y}{2(1+6 \lambda)M_{\nu}} \ee where 
\ba X &=& -3 \vec{S}(M_{\nu}=0)  + 4\vec{S}(M_{\nu}) - \vec{S}(2M_{\nu}); \nonumber \\
      Y &=& -15 \vec{S}(M_{\nu}=0) +16\vec{S}(M_{\nu}) - \vec{S}(4M_{\nu}) \ea and $\lambda$ is a free parameter. The scheme in Eq. ({\ref{eq:Dmnudef}}) corresponds to $\lambda=-1/12$, and the three-term scheme shown in\cite{Quijote} corresponds to $\lambda=0$. One can, in principle, tune $\lambda$ to optimise the convergence of the Fisher matrix. In practice, we numerically tune $\lambda$ for the best performance of the convergence, and this finds $\lambda\sim-1/4$. This effectively removes the $\vec{S}(M_{\nu})$ term, which is understandable - the change in neutrino masses is the smallest in the $\vec{S}(M_{\nu})$ term, compared to that in $\vec{S}(2M_{\nu})$ or $\vec{S}(4M_{\nu})$ terms, thus it may be the one that is most subject to numerical issues. Thus the final expression we use for $\frac{\partial \vec{S}}{\partial M_{\nu}}$ is,
\ba \label{eq:Dmnudef3} \frac{\partial \vec{S}}{\partial M_{\nu}} &\sim& \frac{-\vec{S}(4M_{\nu})  + 4 \vec{S}(2M_{\nu})- 3 \vec{S}(M_{\nu}=0) }{4 M_{\nu} }\bigg|_{M_{\nu} = 0.1 \ {\rm eV}} \nonumber \\ 
&=&  \frac{-\vec{S}(2M_{\nu})  + 4 \vec{S}(M_{\nu})- 3 \vec{S}(M_{\nu}=0) }{2 M_{\nu} }\bigg|_{M_{\nu} = 0.2 \ {\rm eV}}
\ea
The convergence of $\sigma_\theta$ against number of mocks is illustrated in panels $\bf b$ and $\bf c$ in Supplement Figure 5 using two different estimators for the derivative of $M_{\nu}$: panel $\bf b$ uses Eq.~(\ref{eq:Dmnudef}) while $\bf c$ uses Eq.~(\ref{eq:Dmnudef3}). As shown, the convergence is better using our improved estimator (panel $\bf c$) than the one used in the literature (panel $\bf b$), and it shows that our result is sufficiently stable when $N_{\rm mock}>5500$, demonstrating the robustness of our result. We therefore adopt Eq.~(\ref{eq:Dmnudef3}) for all the relevant calculation in this work. 

\newpage

\begin{figure}
\begin{center}
\includegraphics[width=\linewidth]{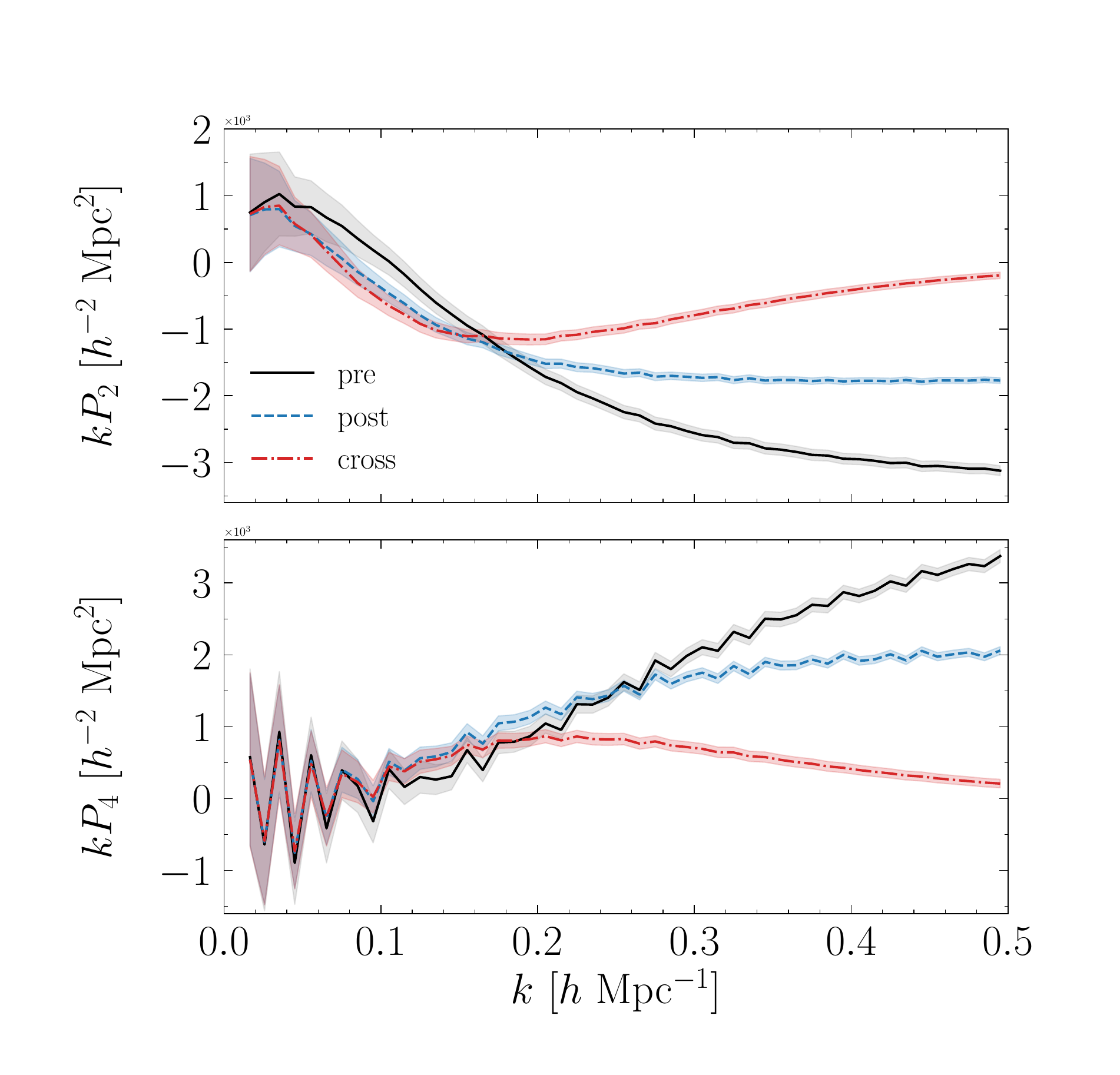}
\end{center}
\end{figure} 

\noindent{\tp{\bf Supplement Figure 1. The quadrupole and hexadecapole moments measured from the Molino catalog.}}

\clearpage

\begin{figure}
\begin{center}
\includegraphics[width=0.9\linewidth]{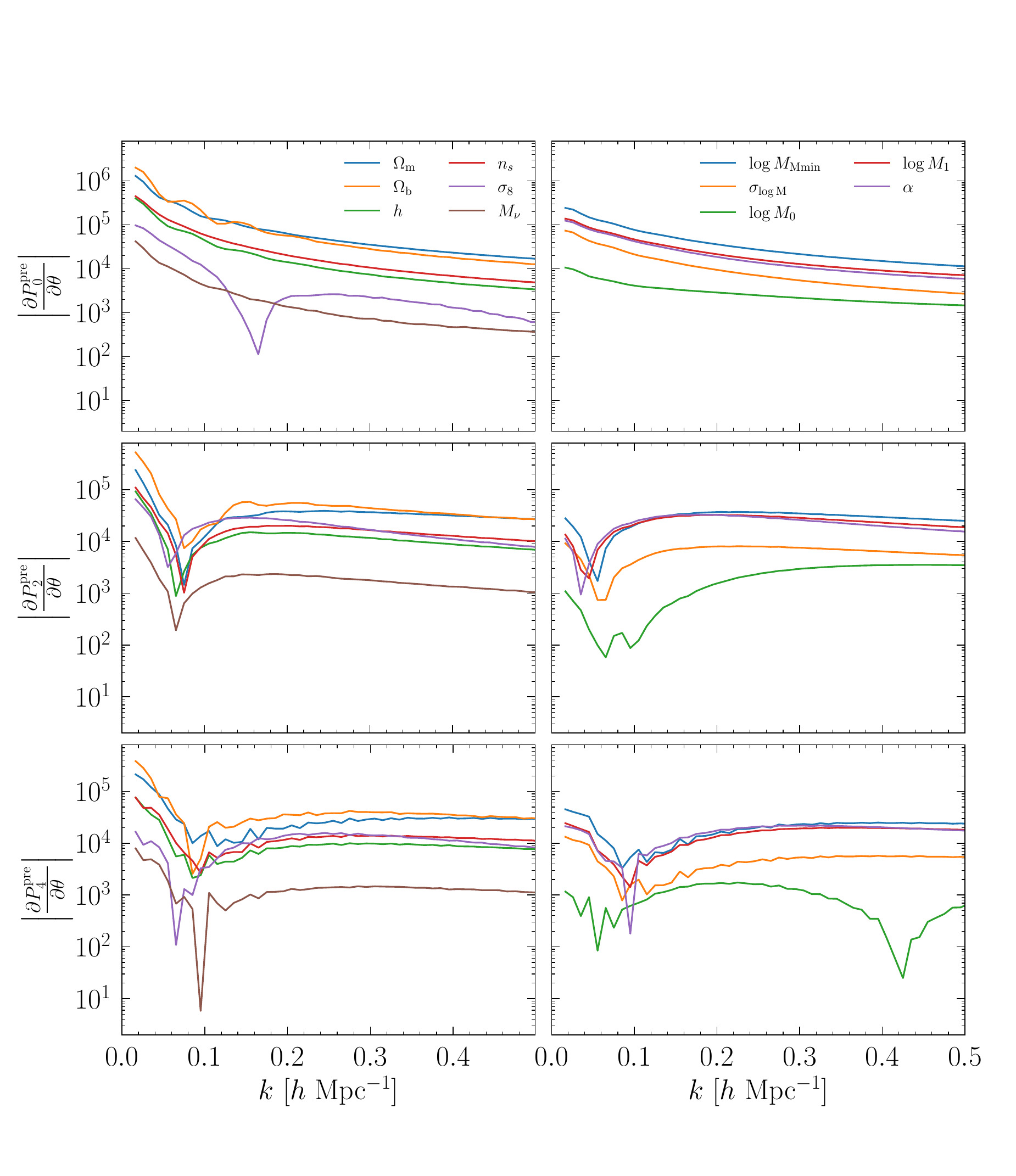}
\end{center}
\end{figure} 

\noindent{\tp{\bf Supplement Figure 2. The derivative of $P_{\rm pre}$ with respect to cosmological parameters and HOD parameters.}}

\clearpage

\begin{figure}
\begin{center}
\includegraphics[width=0.9\linewidth]{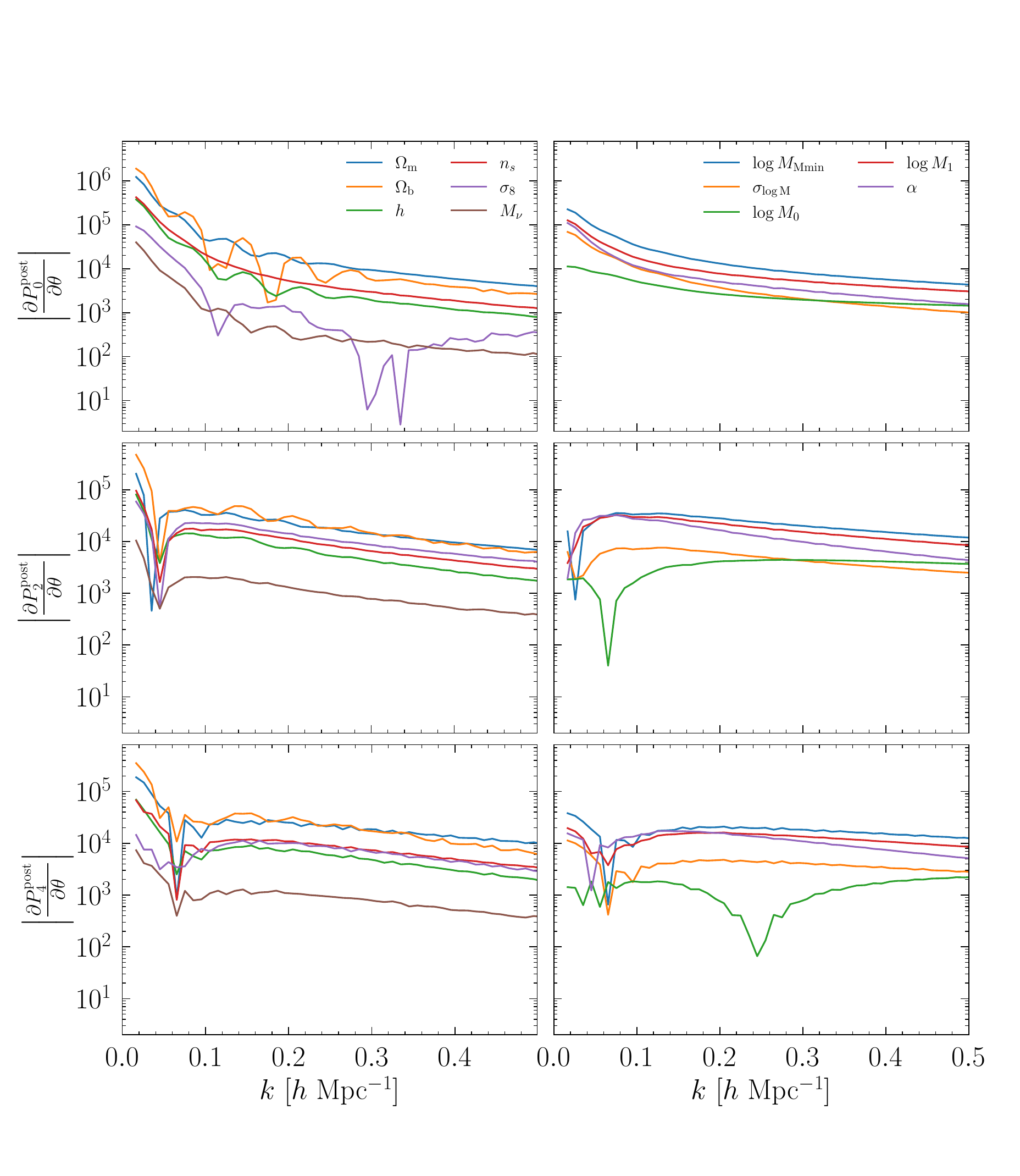}
\end{center}
\end{figure} 

\noindent{\tp{\bf Supplement Figure 3. The derivative of $P_{\rm post}$ with respect to cosmological parameters and HOD parameters.}}

\clearpage

\begin{figure}
\begin{center}
\includegraphics[width=0.9\linewidth]{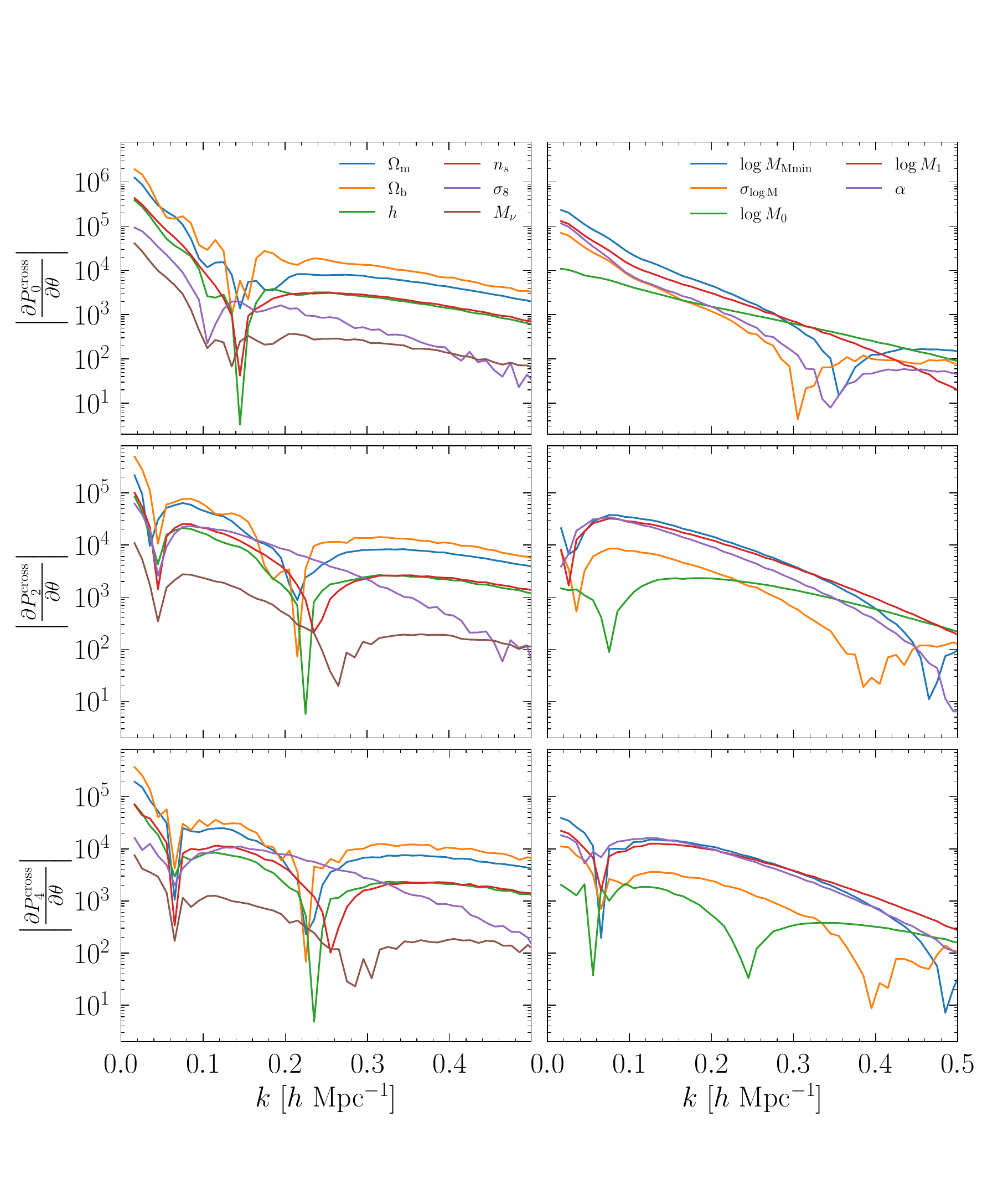}
\end{center}
\end{figure} 

\noindent{\tp{\bf Supplement Figure 4. The derivative of $P_{\rm cross}$ with respect to cosmological parameters and HOD parameters.}}

\clearpage

\begin{figure}
\begin{center}
\includegraphics[width=\linewidth]{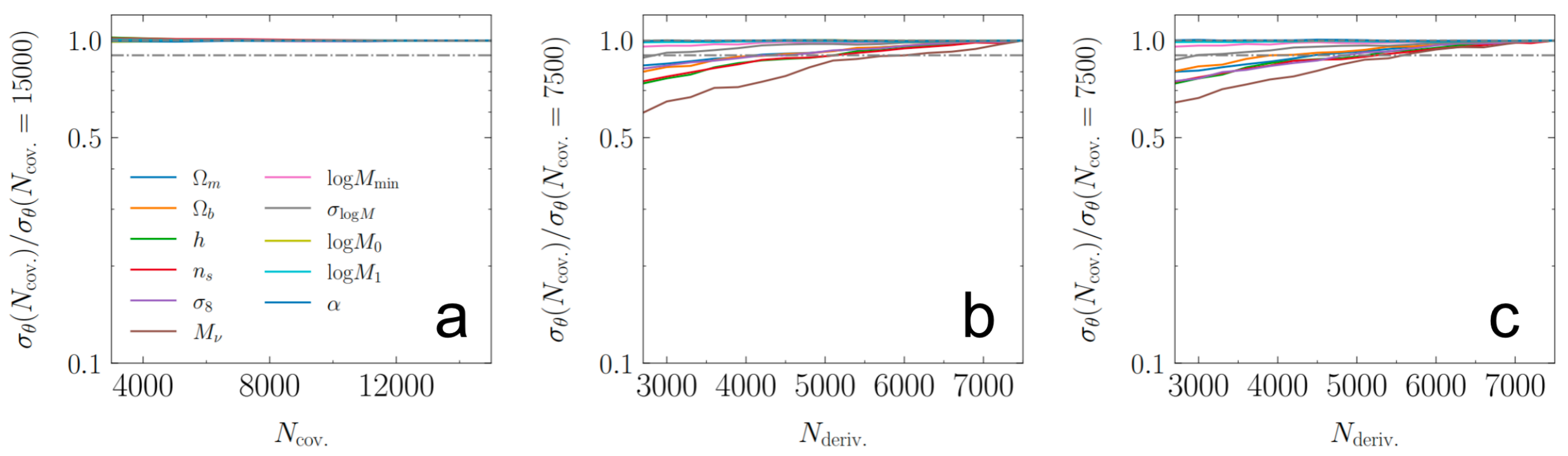}
\end{center}
\end{figure} 

\noindent{\tx{\bf Supplement Figure 5. The convergence of the Fisher result against the number of mocks.}}\\
\tx{Panel {\bf a}: The marginalised uncertainties of cosmological parameters, denoted as $\sigma_\theta$ in general, as a function of number of mocks, which is $15,000$ in total, used to estimate the covariance matrix; Panels {\bf b} and {\bf c}: $\sigma_\theta$ as a function of number of mocks, which is $7,500$ in total, used to estimate the derivatives numerically. In Panel {\bf b}, Eq.~(\ref{eq:Dmnudef}) is used to evaluate the derivative for $M_{\nu}$, while in Panel {\bf c}, the derivative for $M_{\nu}$ is computed using Eq.(~\ref{eq:Dmnudef3}).}

\clearpage

\noindent {\bf Supplementary Note 3. Other supplement figures and tables}

\begin{figure}
\begin{center}
\includegraphics[width=\linewidth]{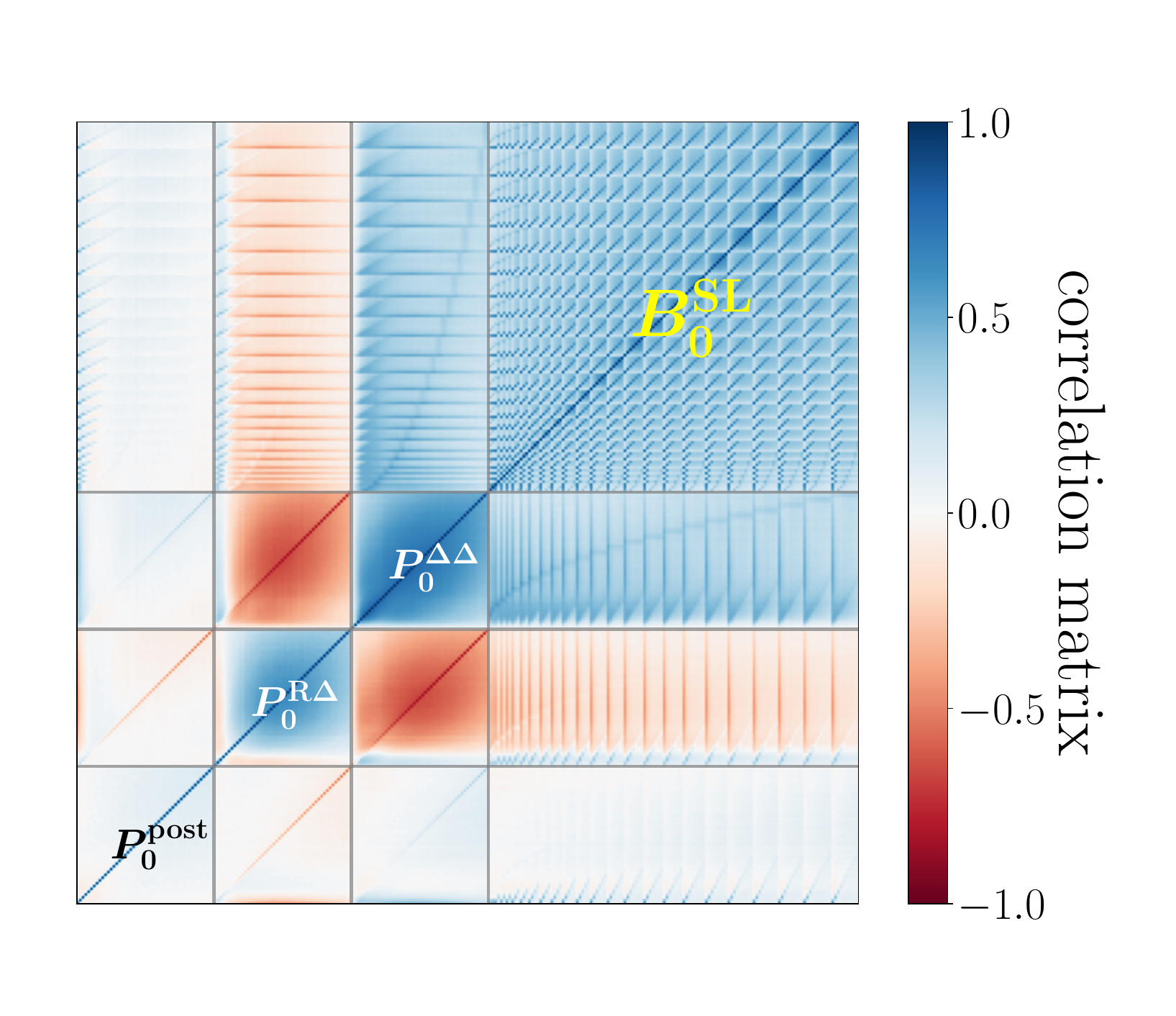}
\end{center}
\end{figure} 

\noindent \tx{\bf Supplement Figure 6. Part of the correlation matrix between the power spectra and bispectrum.}\\
The correlation matrix for the monopole of three types of power spectra ($P_0^{\rm post}, P_0^{R\Delta}, P_0^{\Delta\Delta}$), and of the bispectrum in the squeezed limit ($B_0^{\rm SL}$), \ie\,$k_1=k_2 \gg k_3$, derived from the {\sc Molino} galaxy mocks. The horizontal and vertical lines separate each block for visualisation. For all blocks, the associated $k$ or $k_1$ increases from $0.01$ to $0.5~h~{\rm Mpc}^{-1}$, from left to right, and from bottom to top.

\clearpage

\begin{table}
\begin{adjustbox}{width=\columnwidth,center}
\begin{tabular}{c|cccccc|cccccc}

\hline\hline
                                            & \multicolumn{6}{c|}{$k_{\rm max}=0.2~h~{\rm Mpc}^{-1}$} & \multicolumn{6}{c}{$k_{\rm max}=0.5~h~{\rm Mpc}^{-1}$} \\
 \cline{2-13}  
                                            & \multicolumn{2}{c|}{${\rm w/o\,AP} $} & \multicolumn{2}{c|}{\tc{$z=0.5\,{\rm AP}$}} & \multicolumn{2}{c|}{\tc{$z=1.0\,{\rm AP}$}}  & \multicolumn{2}{c|}{${\rm w/o\,AP}$} & \multicolumn{2}{c|}{\tc{$z=0.5\,{\rm AP}$}} & \multicolumn{2}{c}{\tc{$z=1.0\,{\rm AP} $}}   \\ 
\cline{2-13}                         
                                           & $P_{\rm all}$  & \multicolumn{1}{c|}{$P_{\rm pre} + B_0$}   & $P_{\rm all}$  & \multicolumn{1}{c|}{$P_{\rm pre} + B_0$}   & $P_{\rm all}$  & \multicolumn{1}{c|}{$P_{\rm pre} + B_0$}   & $P_{\rm all}$  & \multicolumn{1}{c|}{$P_{\rm pre} + B_0$}  & $P_{\rm all}$  & \multicolumn{1}{c|}{$P_{\rm pre} + B_0$}   & $P_{\rm all}$  & $P_{\rm pre} + B_0$ \\                
                                                          
\hline
$\sigma(\Omega_{\rm m})$ & $1$&$1.07$ &$0.56$ &$0.70$ &$0.36$ &$0.51$& $1$ &$0.86$ &$0.32$&$0.29$ &$0.21$ &$0.21$\\
$\sigma(\Omega_{\rm b})$ &$1$&$1.17$ &$0.89$ &$1.12$ &$0.84$&$1.10$& $1$ &$0.76$&$0.98$&$0.76$&$0.97$&$0.76$\\
$\sigma(h)$ &$1$&$1.09$ &$0.82$ &$0.98$&$0.78$ &$0.96$ & $1$&$0.64$&$0.94$&$0.64$&$0.94$&$0.64$\\
$\sigma(n_{\rm s})$ & $1$&$1.09$ &$0.82$&$0.90$ &$0.76$&$0.85$& $1$ &$0.63$&$0.82$&$0.55$&$0.81$&$0.55$\\
$\sigma(\sigma_8)$ &$1$ &$1.00$ &$0.89$&$1.00$&$0.81$&$0.93$& $1$&$0.73$&$0.66$&$0.64$&$0.62$&$0.64$\\
$\sigma(M_{\nu}~[{\rm eV}])$ &$1$&$0.96$ &$0.83$&$0.84$&$0.81$&$0.83$& $1$ &$0.47$&$0.89$&$0.46$&$0.89$&$0.46$\\\hline
$\sigma({\rm log}M_{\rm min})$ &$1$&$1.77$&$1.00$&$1.79$&$1.00$&$1.80$& $1$ &$2.07$&$0.95$&$2.04$&$0.94$&$2.08$\\
$\sigma(\sigma_{\rm log}$$_{M})$ &$1$&$2.44$&$1.05$&$2.45$ &$1.04$&$2.45$& $1$ &$1.80$&$1.04$&$1.78$&$1.03$&$1.78$\\
$\sigma({\rm log}M_{0})$ &$1$&$3.09$&$1.00$&$3.06$ &$1.00$&$3.05$& $1$ &$1.29$&$1.03$&$1.32$&$1.03$&$1.32$\\
$\sigma({\rm log}M_{1})$ &$1$&$1.15$&$1.00$&$1.14$ &$1.00$ &$1.15$& $1$ &$1.20$&$1.00$&$1.19$&$1.00$&$1.22$\\
$\sigma(\alpha)$ &$1$&$1.94$&$1.00$&$1.90$&$1.00$ &$1.90$& $1$ &$0.93$&$1.01$&$0.89$&$1.01$&$0.89$\\\hline
\hline

\end{tabular}
\end{adjustbox}
\\\\
\tx{\bf Supplement Table 1: Uncertainties on parameters derived from the {\sc Molino} mock with various implications of the AP effect.} \\
The 68\% confidence level constraints on parameters derived from the {\sc Molino} mock at $z=0$ without and with the AP effect implemented using pairs of ($\alpha_{\bot}, \alpha_{||}$) calculated at $z=0.5$ and $1$, respectively. All quantities are normalised by the result of $P_{\rm all}$ without the AP effect.        
\end{table}%

\clearpage

\begin{figure}
\begin{center}
\includegraphics[width=\linewidth]{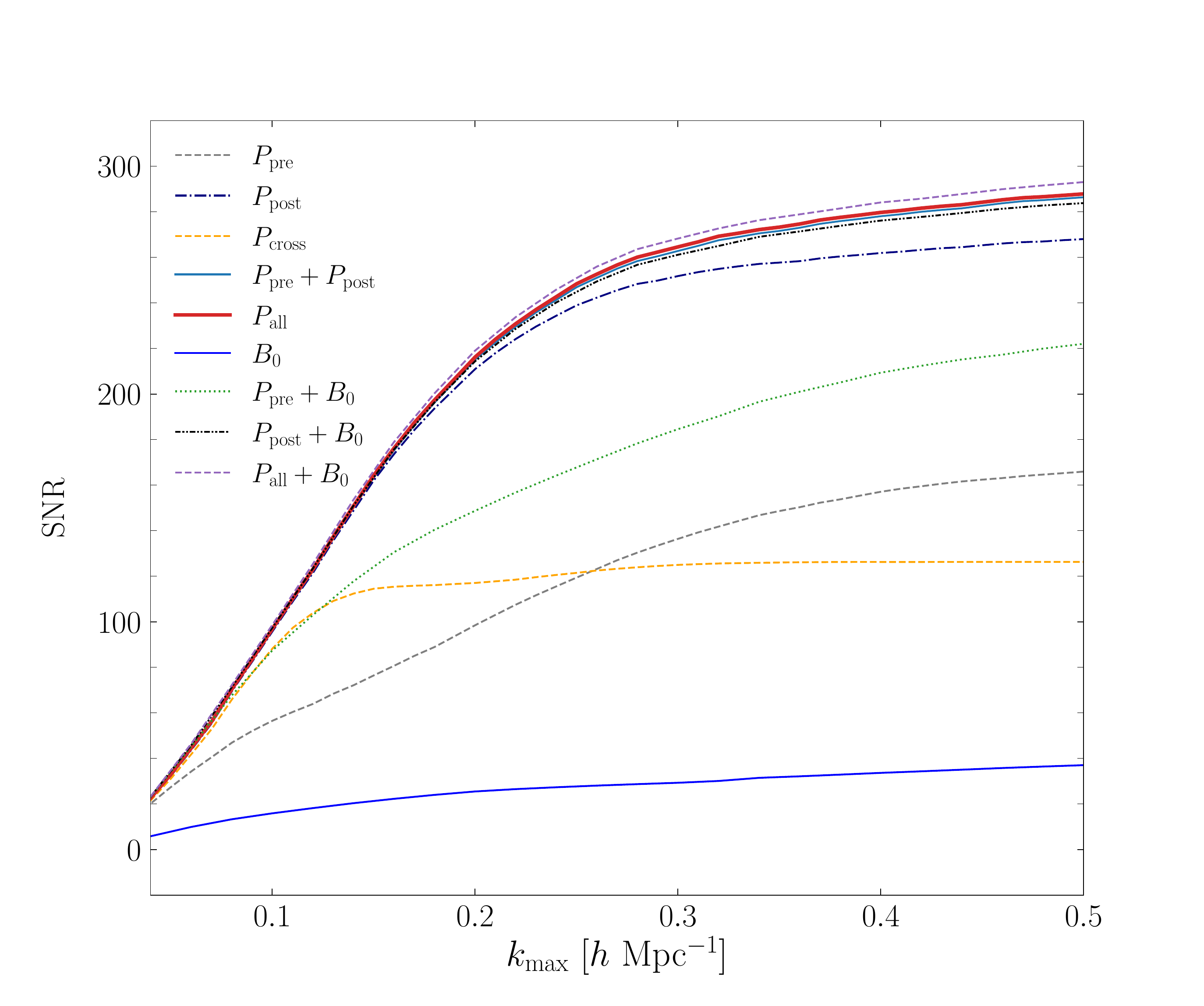}
\end{center}
\end{figure} 

\noindent{\tc{\bf Supplement Figure 7. The SNR of the spectra measured from the {\sc Molino} mock}.\\
\tc{The cumulative signal-to-noise ratio for various datasets as a function of the maximum wave number.}

\begin{figure}
\begin{center}
\includegraphics[width=\linewidth]{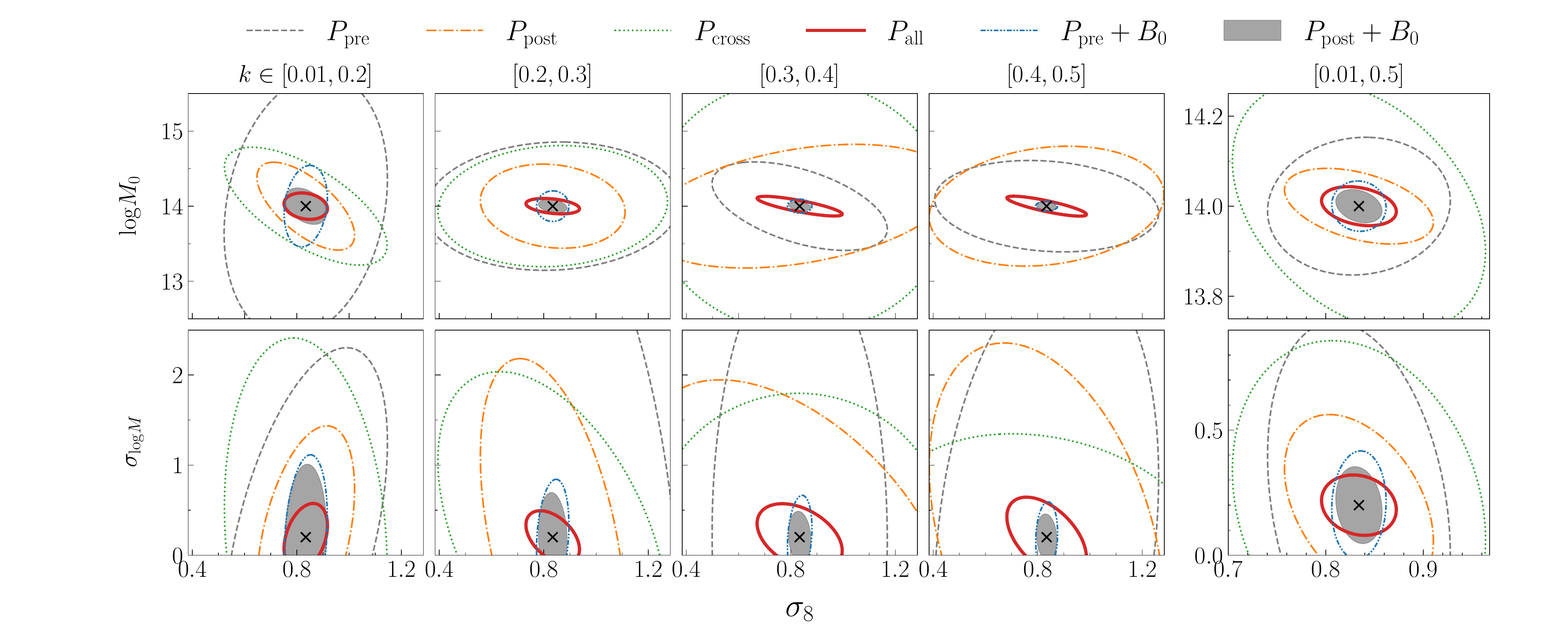}
\end{center}
\end{figure} 

\noindent \tc{\bf Supplement Figure 8. Contour plots derived from the {\sc Molino} mock.} \\
\tc{The 68\% CL contour plots for ($\sigma_8$, ${\rm log}M_0$) and ($\sigma_8$, $\sigma_{\rm log}$$_M$) derived from various combinations of spectra measured from the {\sc Molino} mock in several $k$ ranges.}

\clearpage

\begin{figure}
\begin{center}
\includegraphics[width=\linewidth]{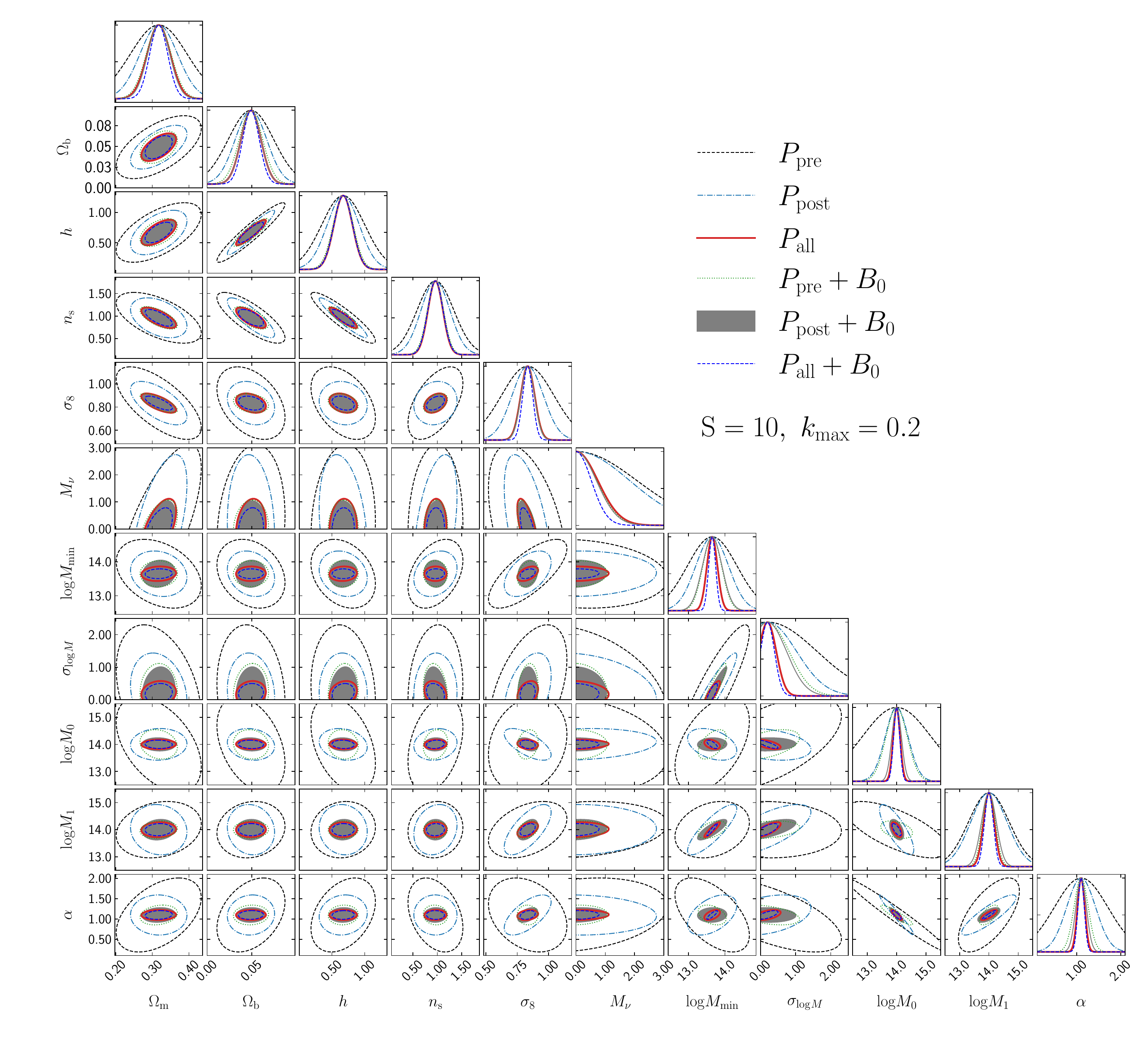}
\end{center}
\end{figure} 

\noindent \tc{\bf Supplement Figure 9. Contour plots derived from the {\sc Molino} mock.} \\
\tc{The 1D posterior distribution and 68\% CL contour plots for cosmological and HOD parameters measured from the {\sc Molino} mock with $k_{\rm max}=0.2~h~\rm Mpc^{-1}$. The smoothing scale for the reconstruction is $S=10~h^{-1}~{\rm Mpc}$}.

\clearpage

\begin{figure}
\begin{center}
\includegraphics[width=\linewidth]{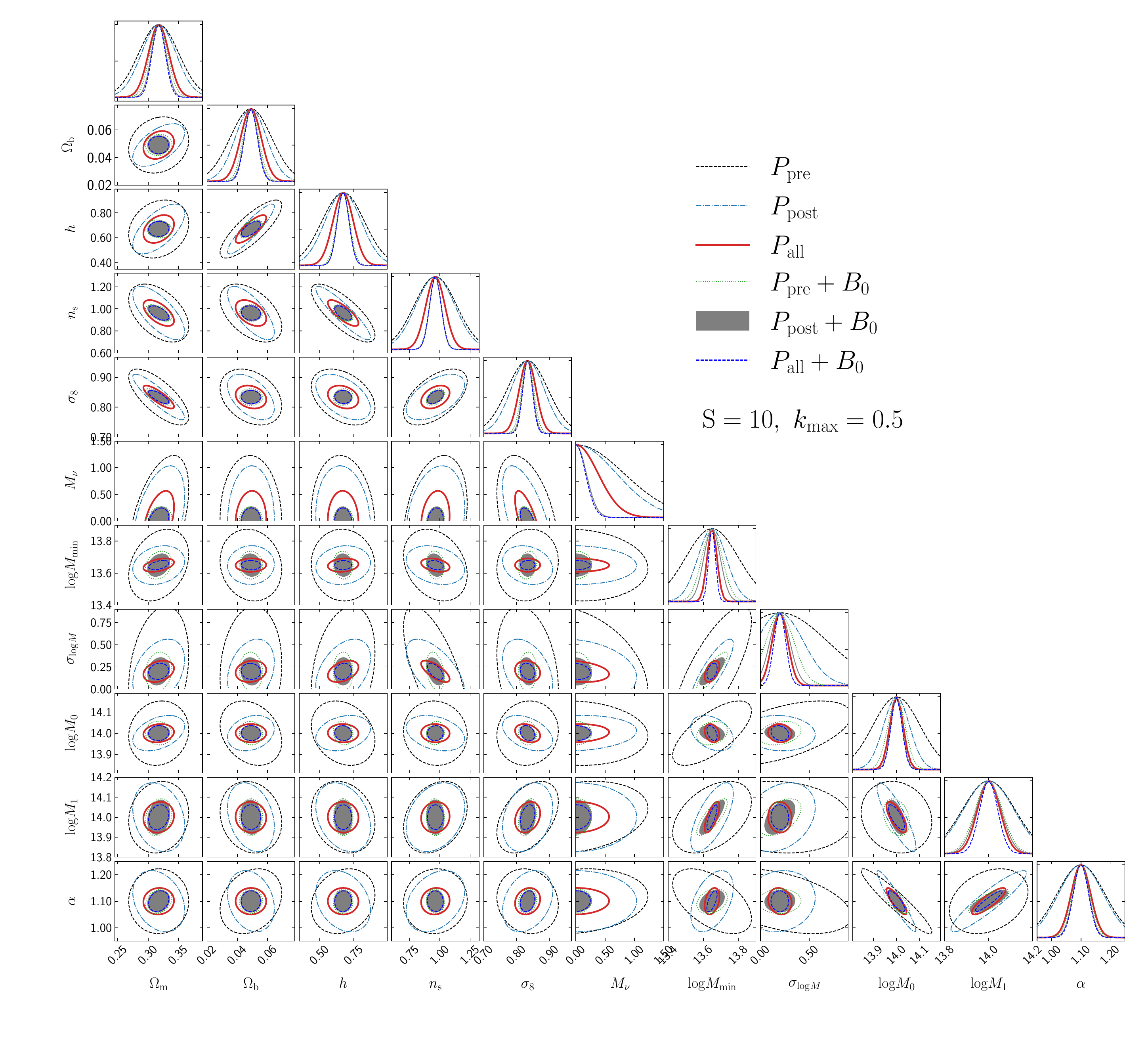}
\end{center}
\end{figure} 

\noindent \tc{\bf Supplement Figure 10. Contour plots derived from the {\sc Molino} mock.} \\ 
\tc{The 1D posterior distribution and 68\% CL contour plots for cosmological and HOD parameters measured from the {\sc Molino} mock with $k_{\rm max}=0.5~h~\rm Mpc^{-1}$. The smoothing scale for the reconstruction is $S=10~h^{-1}~{\rm Mpc}$}.

\clearpage

\begin{figure}
\begin{center}
\includegraphics[width=\linewidth]{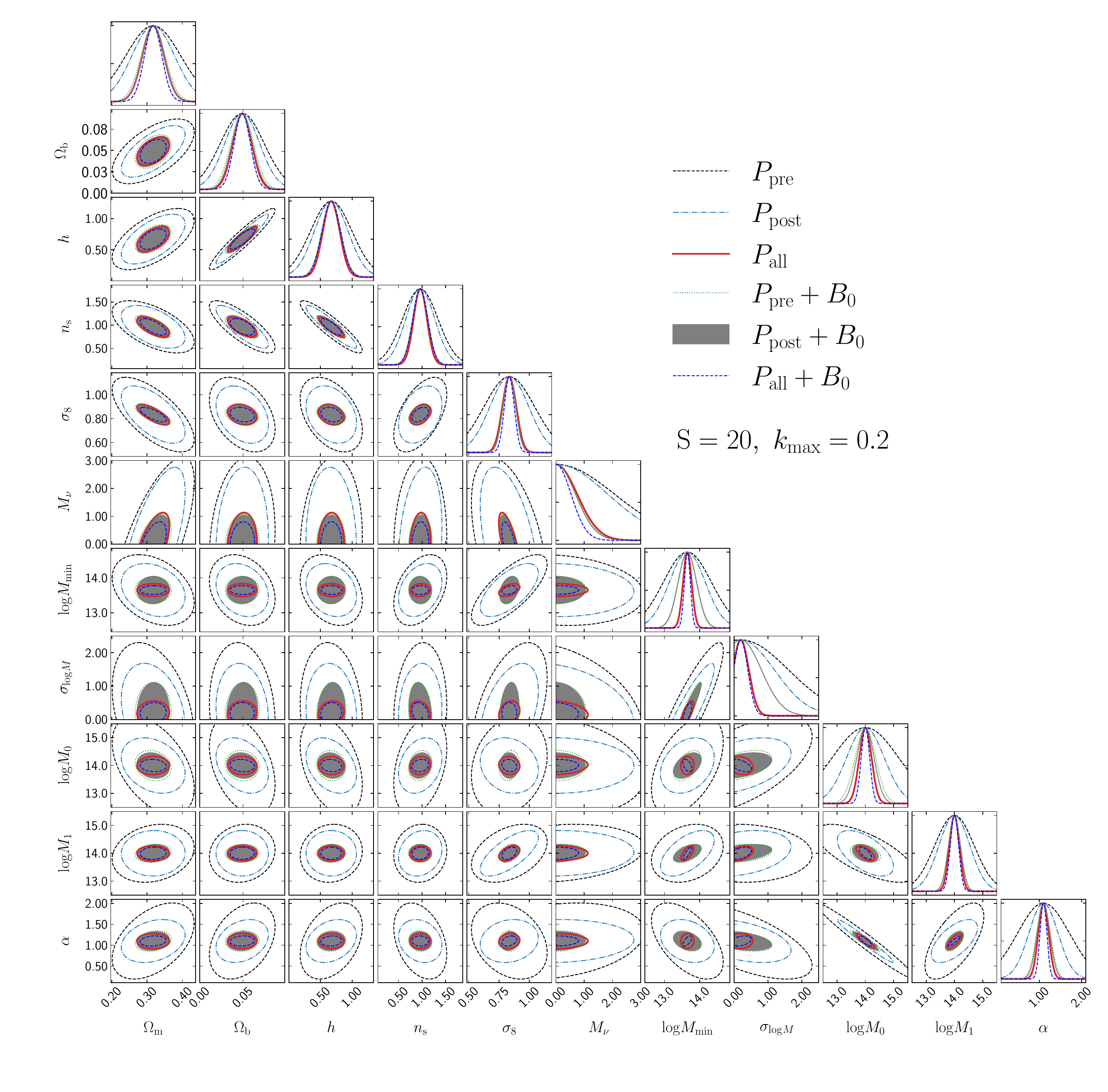}
\end{center}
\end{figure} 

\noindent \tc{\bf Supplement Figure 11. Contour plots derived from the {\sc Molino} mock.} \\
\tc{The 1D posterior distribution and 68\% CL contour plots for cosmological and HOD parameters measured from the {\sc Molino} mock with $k_{\rm max}=0.2~h~\rm Mpc^{-1}$. The smoothing scale for the reconstruction is $S=20~h^{-1}~{\rm Mpc}$}.

\clearpage

\begin{figure}
\begin{center}
\includegraphics[width=\linewidth]{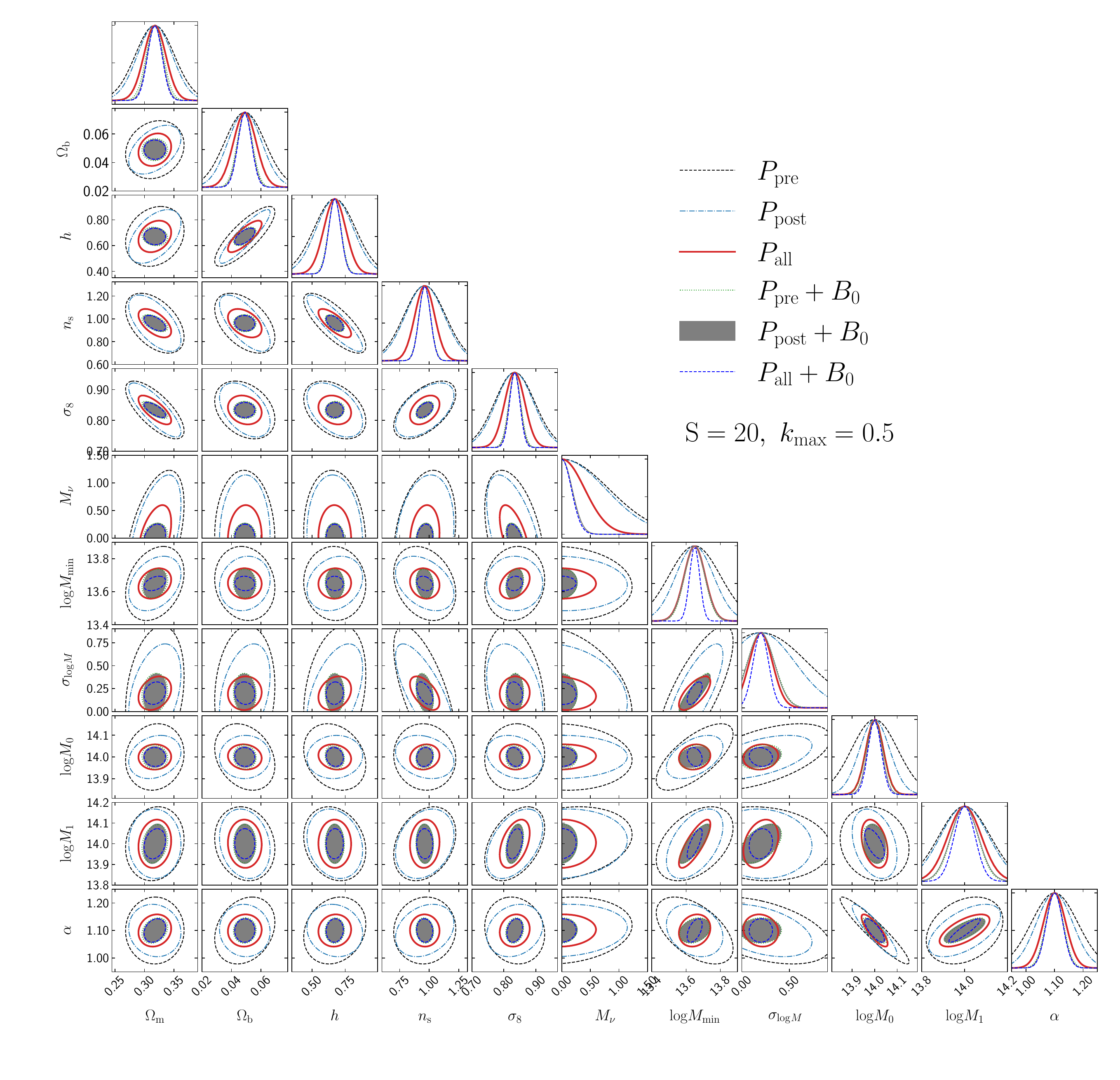}
\end{center}
\end{figure} 

\noindent \tc{\bf Supplement Figure 12. Contour plots derived from the {\sc Molino} mock.} \\
\tc{The 1D posterior distribution and 68\% CL contour plots for cosmological and HOD parameters measured from the {\sc Molino} mock with $k_{\rm max}=0.5~h~\rm Mpc^{-1}$. The smoothing scale for the reconstruction is $S=20~h^{-1}~{\rm Mpc}$}.

\clearpage 

\begin{figure}
\begin{center}
\includegraphics[width=\linewidth]{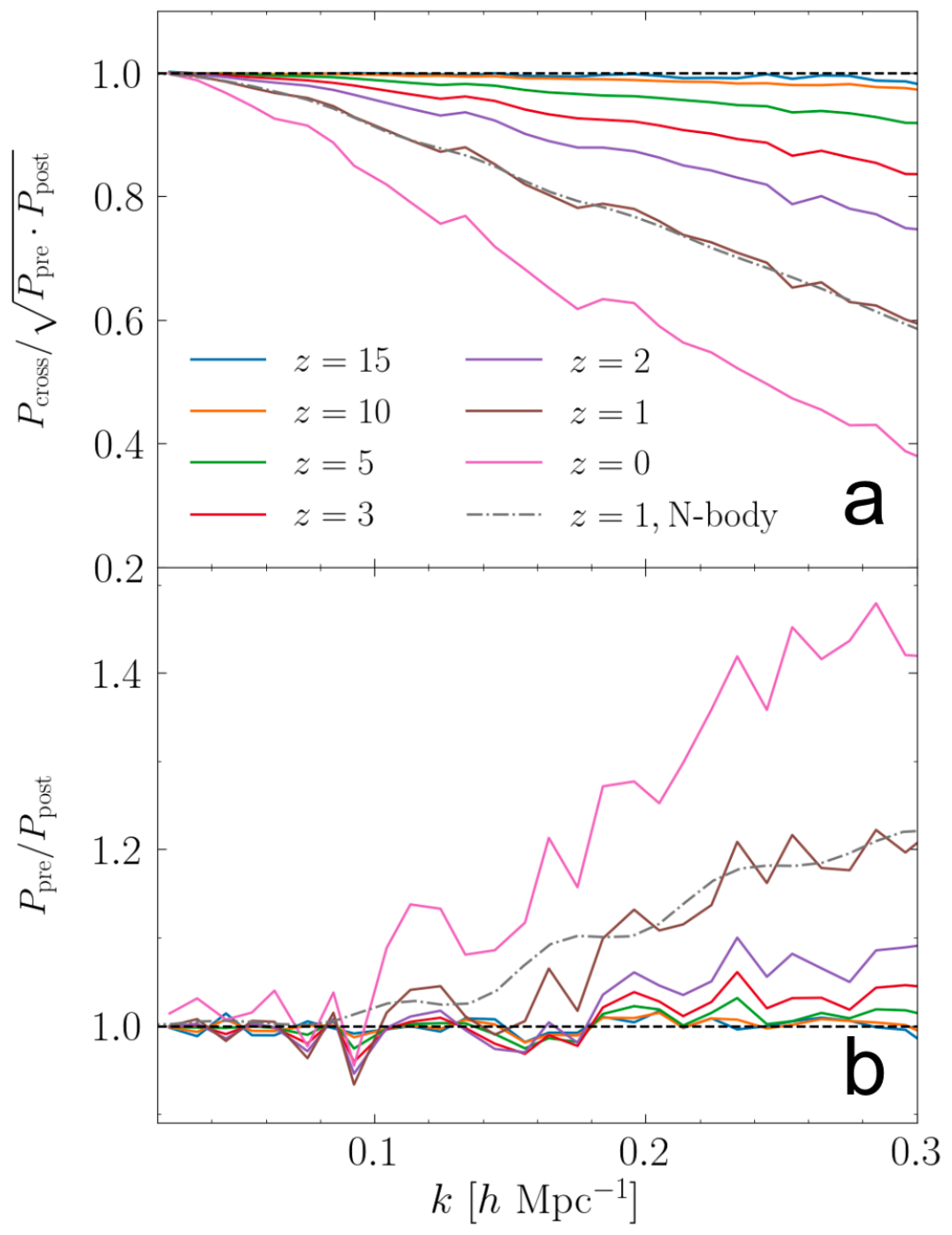}
\end{center}
\end{figure} 

\noindent \tx{\bf Supplement Figure 13. The decorrelation effect as a function of scales at various redshifts} \\
\tx{Panel {\bf a}: Solid: the decorrelation between the monopole of $P_{\rm pre}$ and $P_{\rm post}$, as quantified by $P_{\rm cross}/\sqrt{P_{\rm pre} \cdot P_{\rm post}}$, as a function of redshift $z$ derived from the COLA mocks; Dash-dotted: the decorrelation derived from the $N$-body mocks at $z=1$. The horizontal dash line shows the perfect correlation between $P_{\rm pre}$ and $P_{\rm post}$ as a reference; Panel {\bf b}: The ratio between $P_{\rm pre}$ and $P_{\rm post}$ at various redshifts. The horizontal dash line shows the prediction of the linear perturbation theory.}

\clearpage

\begin{figure}
\begin{center}
\includegraphics[width=\linewidth]{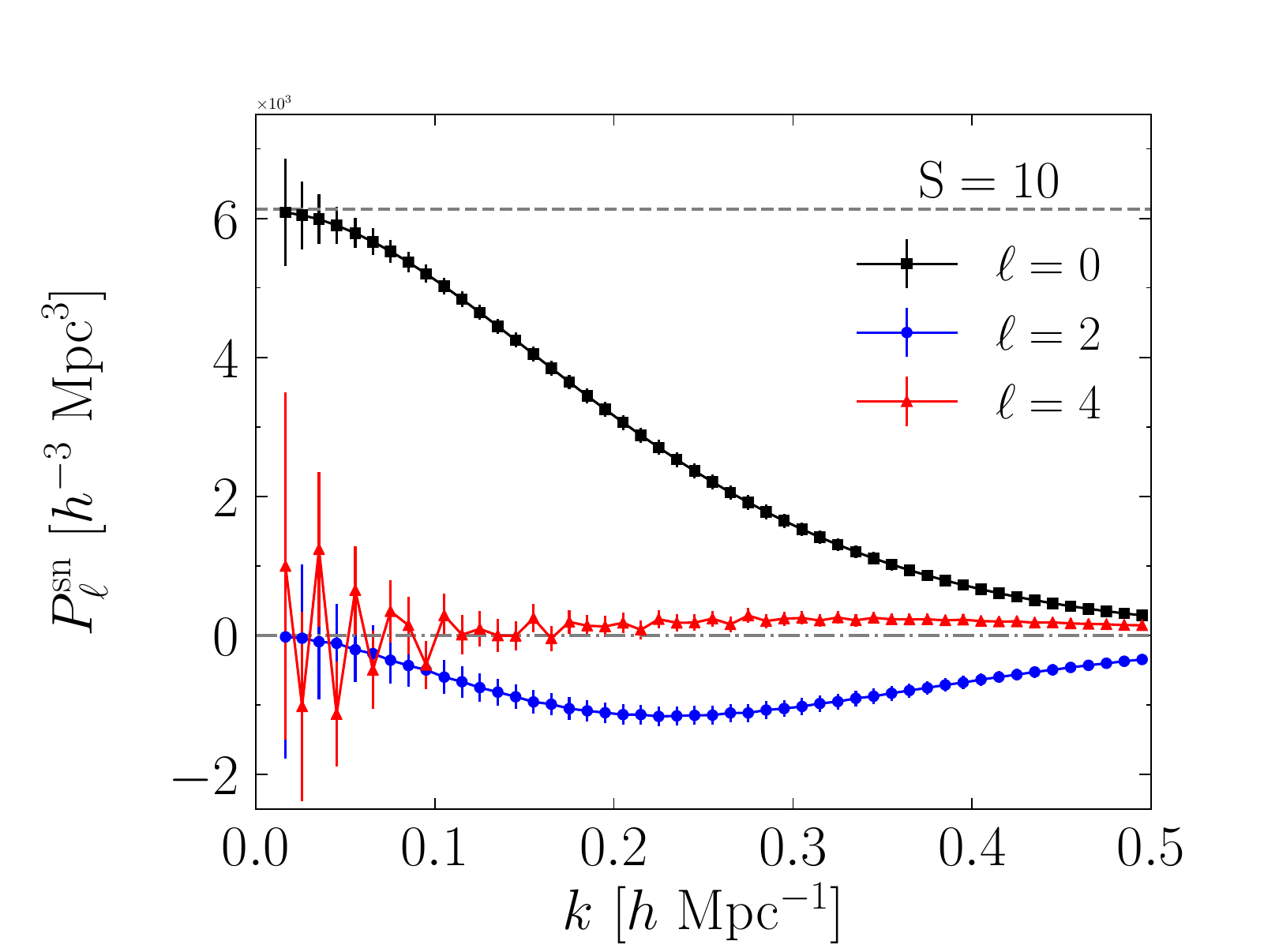}
\end{center}
\end{figure} 

\noindent \tc{\bf Supplement Figure 14. The noise power spectrum of the cross power spectrum.} \\
\tc{The noise power spectrum for the cross power spectrum multipoles measured from the {\sc Molino} mock. The smoothing scale $S=10~h^{-1}~{\rm Mpc}$ is used for the reconstruction.}

\clearpage

\begin{figure}
\begin{center}
\includegraphics[width=\linewidth]{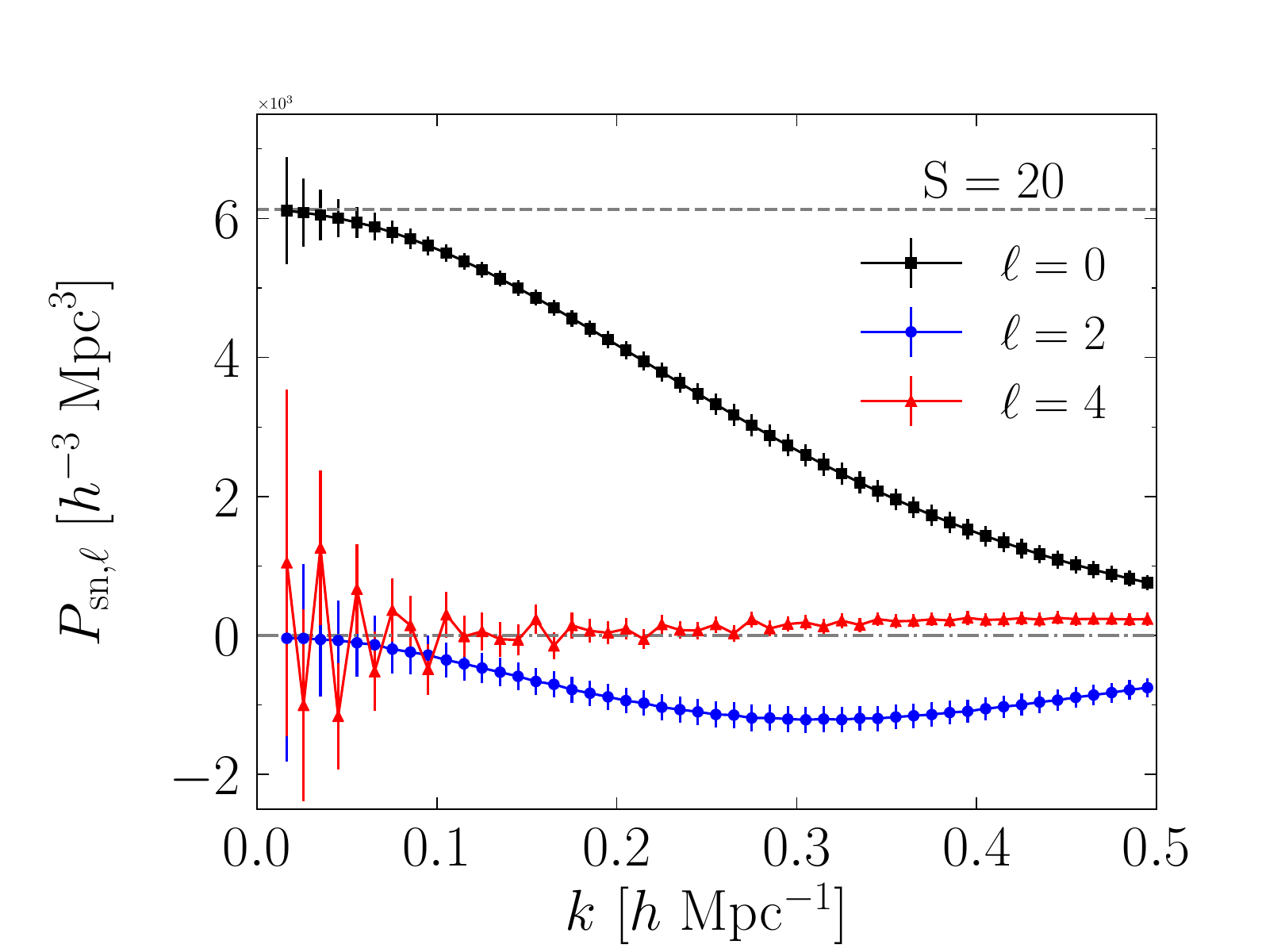}
\end{center}
\end{figure} 

\noindent \tc{\bf Supplement Figure 15. The noise power spectrum of the cross power spectrum.} \\
\tc{The noise power spectrum for the cross power spectrum multipoles measured from the {\sc Molino} mock. The smoothing scale $S=20~h^{-1}~{\rm Mpc}$ is used for the reconstruction.}

\begin{figure}
\begin{center}
\includegraphics[width=\linewidth]{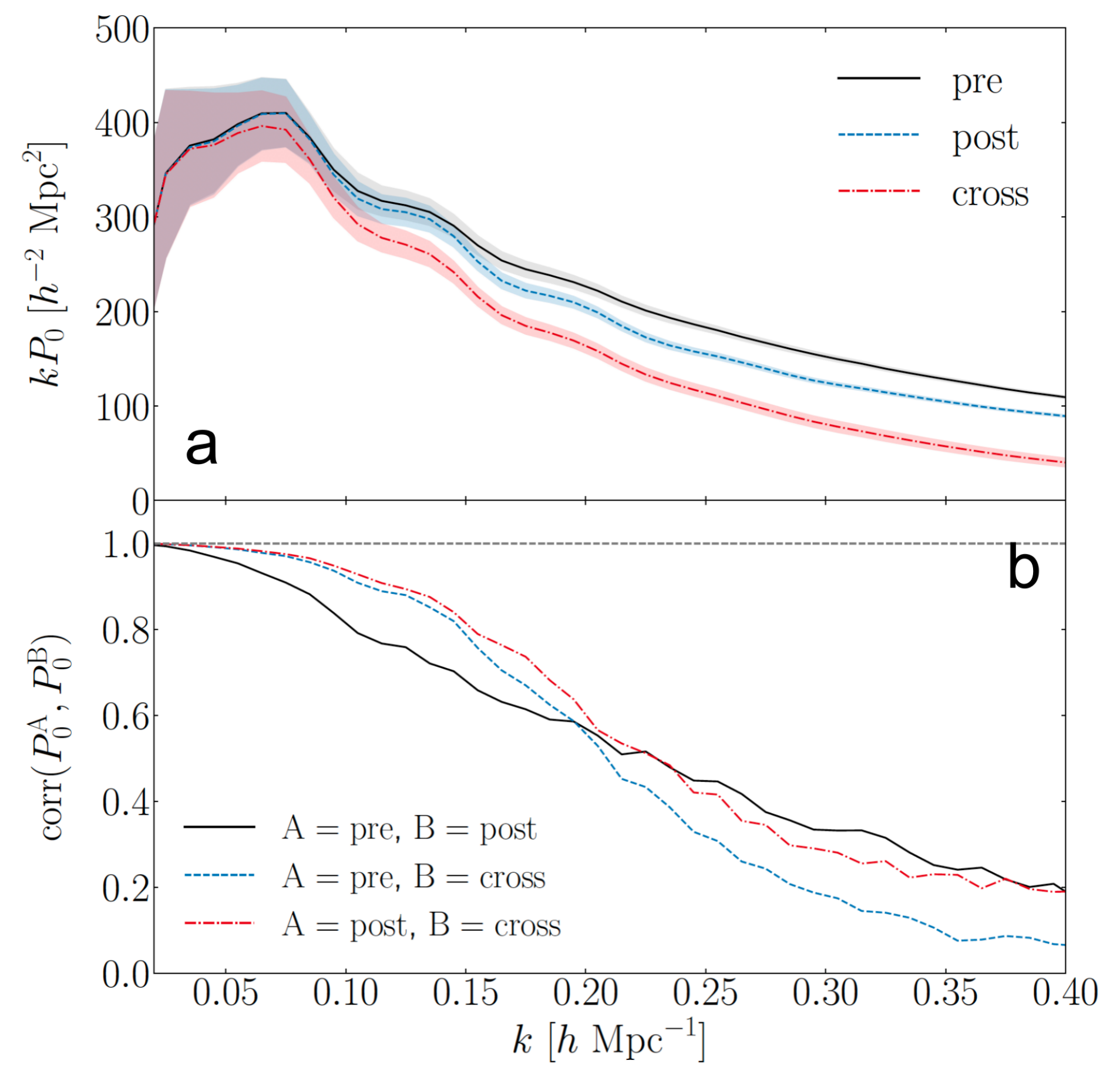}
\end{center}
\end{figure} 

\noindent {\bf Supplement Figure 16. The measured three types of power spectra and correlation among them.}  \\
Panel {\bf a}: The mean (lines) and 68\% uncertainty (shades) of three types of power spectrum monopole (multiplied by the wavenumber $k$) measured from the $4000$ $N$-body mocks at $z=1.02$;  Panel {\bf b}: The measured correlation coefficient between various types of power spectrum monopole. The dashed horizontal line shows a perfect correlation (\ie, ${\rm corr}=1$) for a reference.

\clearpage

\begin{figure}
\begin{center}
\includegraphics[width=0.9\linewidth]{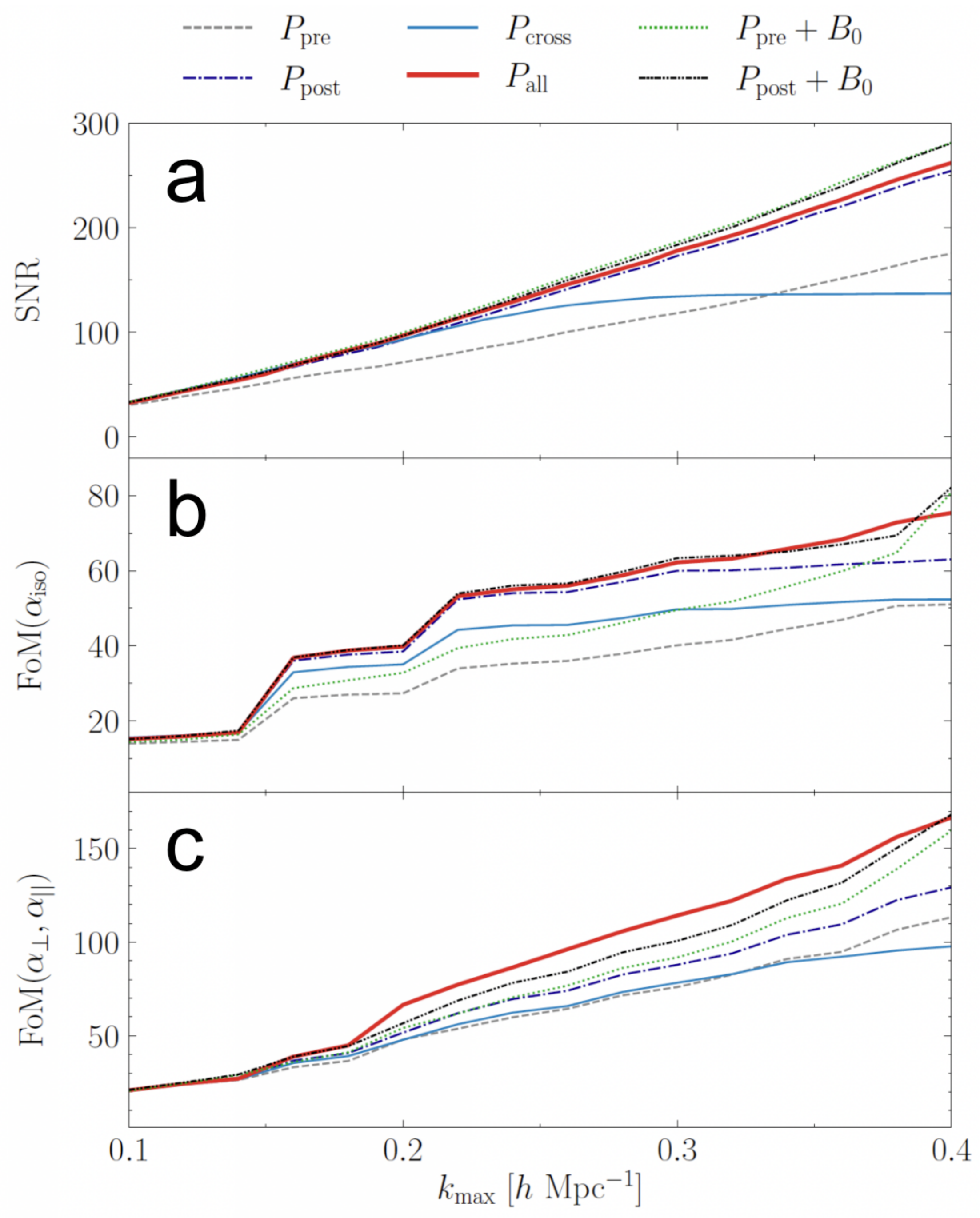}
\end{center}
\end{figure} 

\noindent \tc{\bf Supplement Figure 17. The SNR of the raw measurements and the FoM of the AP parameters.} \\
\tc{Panel {\bf a}: The cumulative signal-to-noise ratio for the measured spectra denoted in the legend, as a function of the maximum wavenumber; Panel {\bf b}: The Figure of Merit of the isotropic dilation parameter \tx{$\alpha_{\rm iso}$} with the overall amplitudes of power spectra and bispectrum monopole marginalised over; Panel {\bf c}: The Figure of Merit of the anisotropic dilation parameters ($\alpha_{\perp}$,$\alpha_{||}$) with the overall amplitudes of power spectrum multipole and bispectrum monopole marginalised over.}

\clearpage

\begin{figure}
\begin{center}
\includegraphics[width=\linewidth]{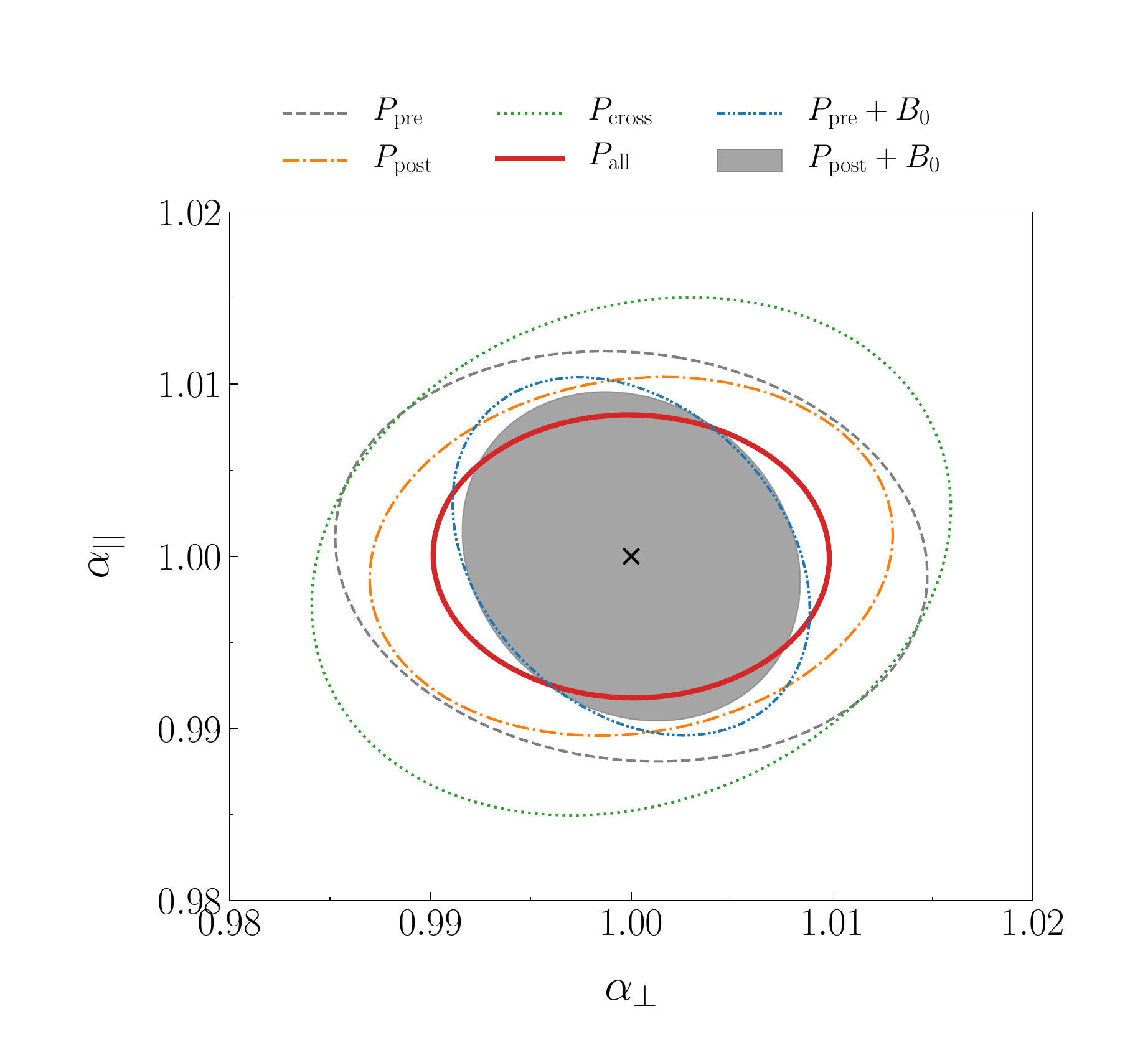}
\end{center}
\end{figure} 

\noindent \tc{\bf Supplement Figure 18. The 68\% CL contour plot for the AP parameters.} \\
\tc{The 68\% CL contour plot for $\alpha_{\perp}$ and $\alpha_{||}$ derived from various data combinations with $k_{\rm max} = 0.4~h~\rm Mpc^{-1}$.}

\clearpage

\begin{figure}
\begin{center}
\includegraphics[width=\linewidth]{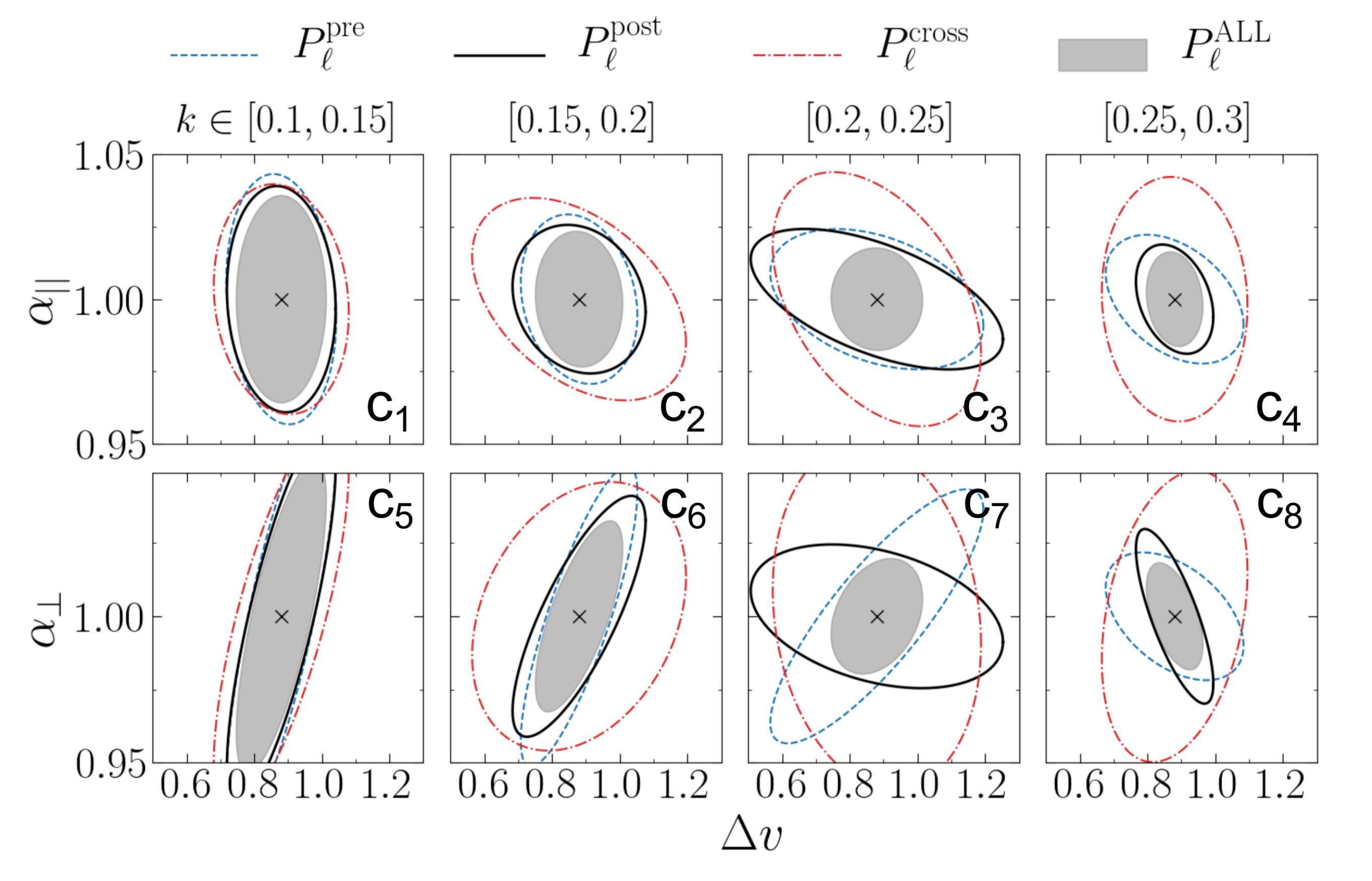}
\end{center}
\end{figure} 

\noindent \tc{\bf Supplement Figure 19. The 68\% CL contour plots for BAO and RSD parameters.} \\
\tc{The 68\% CL contour plots for BAO and RSD parameters including $\alpha_{||},\alpha_{\perp}$ and $\Delta v$ derived from three types of power spectrum multipoles ($\ell=0,2,4$) separately (lines) and jointly (filled). Power spectra used in different $k$ ranges are shown in different columns, as illustrated in the legend.}

\clearpage

\noindent {\bf Supplementary References}


\section{References}

\begin{thebibliography}{30}
\expandafter\ifx\csname url\endcsname\relax
  \def\url#1{\texttt{#1}}\fi
\expandafter\ifx\csname urlprefix\endcsname\relax\def\urlprefix{URL }\fi
\providecommand{\bibinfo}[2]{#2}
\providecommand{\eprint}[2][]{\url{#2}}

\bibitem{BAO98}
\bibinfo{author}{{Eisenstein}, D.~J.} \& \bibinfo{author}{{Hu}, W.}
\newblock \bibinfo{title}{{Baryonic Features in the Matter Transfer Function}}.
\newblock \emph{\bibinfo{journal}{Astrophys. J.}}
  \textbf{\bibinfo{volume}{496}}, \bibinfo{pages}{605--614}
  (\bibinfo{year}{1998}).


\bibitem{DE1}
\bibinfo{author}{{Riess}, A.~G.} \emph{et~al.}
\newblock \bibinfo{title}{{Observational Evidence from Supernovae for an
  Accelerating Universe and a Cosmological Constant}}.
\newblock \emph{\bibinfo{journal}{Astron. J.}} \textbf{\bibinfo{volume}{116}},
  \bibinfo{pages}{1009--1038} (\bibinfo{year}{1998}).

\bibitem{DE2}
\bibinfo{author}{{Perlmutter}, S.} \emph{et~al.}
\newblock \bibinfo{title}{{Measurements of {$\Omega$} and {$\Lambda$} from 42
  High-Redshift Supernovae}}.
\newblock \emph{\bibinfo{journal}{Astrophys. J.}}
  \textbf{\bibinfo{volume}{517}}, \bibinfo{pages}{565--586}
  (\bibinfo{year}{1999}).

\bibitem{BOSS}
\bibinfo{author}{{Dawson}, K.~S.} \emph{et~al.}
\newblock \bibinfo{title}{{The Baryon Oscillation Spectroscopic Survey of
  SDSS-III}}.
\newblock \emph{\bibinfo{journal}{Astron. J.}} \textbf{\bibinfo{volume}{145}},
  \bibinfo{pages}{10} (\bibinfo{year}{2013}).

\bibitem{eBOSSDR16}
\bibinfo{author}{Alam, S.} \emph{et~al.}
\newblock \bibinfo{title}{{Completed SDSS-IV extended Baryon Oscillation
  Spectroscopic Survey: Cosmological implications from two decades of
  spectroscopic surveys at the Apache Point Observatory}}.
\newblock \emph{\bibinfo{journal}{Phys. Rev. D}}
  \textbf{\bibinfo{volume}{103}}, \bibinfo{pages}{083533}
  (\bibinfo{year}{2021}).

\bibitem{recon07}
\bibinfo{author}{Eisenstein, D.~J.}, \bibinfo{author}{Seo, H.-j.},
  \bibinfo{author}{Sirko, E.} \& \bibinfo{author}{Spergel, D.}
\newblock \bibinfo{title}{{Improving Cosmological Distance Measurements by Reconstruction of the Baryon Acoustic Peak}}.
\newblock \emph{\bibinfo{journal}{Astrophys. J.}}
  \textbf{\bibinfo{volume}{664}}, \bibinfo{pages}{675--679}
  (\bibinfo{year}{2007}).

\bibitem{recon12}
\bibinfo{author}{{Padmanabhan}, N.} \emph{et~al.}
\newblock \bibinfo{title}{{A 2 per cent distance to $z = 0.35$ by reconstructing
  baryon acoustic oscillations - I. Methods and application to the Sloan
  Digital Sky Survey}}.
\newblock \emph{\bibinfo{journal}{Mon. Not. Roy. Astron. Soc.}} \textbf{\bibinfo{volume}{427}},
  \bibinfo{pages}{2132--2145} (\bibinfo{year}{2012}).

\bibitem{reconLag}
\bibinfo{author}{{Burden}, A.} \emph{et~al.}
\newblock \bibinfo{title}{{Efficient reconstruction of linear baryon acoustic
  oscillations in galaxy surveys}}.
\newblock \emph{\bibinfo{journal}{Mon. Not. Roy. Astron. Soc.}} \textbf{\bibinfo{volume}{445}},
  \bibinfo{pages}{3152--3168} (\bibinfo{year}{2014}).

\bibitem{reconEuler}
\bibinfo{author}{{Schmittfull}, M.}, \bibinfo{author}{{Feng}, Y.},
  \bibinfo{author}{{Beutler}, F.}, \bibinfo{author}{{Sherwin}, B.} \&
  \bibinfo{author}{{Chu}, M.~Y.}
\newblock \bibinfo{title}{{Eulerian BAO reconstructions and N -point
  statistics}}.
\newblock \emph{\bibinfo{journal}{Phys. Rev. D}} \textbf{\bibinfo{volume}{92}},
  \bibinfo{pages}{123522} (\bibinfo{year}{2015}).

\bibitem{Kaiser}
\bibinfo{author}{Kaiser, N.}
\newblock \bibinfo{title}{{Clustering in real space and in redshift space}}.
\newblock \emph{\bibinfo{journal}{Mon. Not. Roy. Astron. Soc.}}
  \textbf{\bibinfo{volume}{227}}, \bibinfo{pages}{1--27}
  (\bibinfo{year}{1987}).
  
\bibitem{RSD1}
\bibinfo{author}{Lue, A.}, \bibinfo{author}{Scoccimarro, R.} \&
  \bibinfo{author}{Starkman, G.}
\newblock \bibinfo{title}{Differentiating between modified gravity and dark
  energy}.
\newblock \emph{\bibinfo{journal}{Phys. Rev. D}} \textbf{\bibinfo{volume}{69}},
  \bibinfo{pages}{044005} (\bibinfo{year}{2004}).
  
\bibitem{RSD2}
\bibinfo{author}{Guzzo, L.} \emph{et~al.}
\newblock \bibinfo{title}{{A test of the nature of cosmic acceleration using
  galaxy redshift distortions}}.
\newblock \emph{\bibinfo{journal}{Nature}} \textbf{\bibinfo{volume}{451}},
  \bibinfo{pages}{541--545} (\bibinfo{year}{2008}).


\bibitem{Hikage2020}
\bibinfo{author}{Hikage, C.}, \bibinfo{author}{Takahashi, R.} \&
  \bibinfo{author}{Koyama, K.}
\newblock \bibinfo{title}{{Covariance of the redshift-space matter power
  spectrum after reconstruction}}.
\newblock \emph{\bibinfo{journal}{Phys. Rev. D}}
  \textbf{\bibinfo{volume}{102}}, \bibinfo{pages}{083514}
  (\bibinfo{year}{2020}).

\bibitem{Zhu18}
\bibinfo{author}{{Zhu}, H.-M.}, \bibinfo{author}{{Yu}, Y.} \&
  \bibinfo{author}{{Pen}, U.-L.}
\newblock \bibinfo{title}{{Nonlinear reconstruction of redshift space
  distortions}}.
\newblock \emph{\bibinfo{journal}{Phys. Rev. D}} \textbf{\bibinfo{volume}{97}},
  \bibinfo{pages}{043502} (\bibinfo{year}{2018}).

\bibitem{Seo15}
\bibinfo{author}{Seo, H.-J.}, \bibinfo{author}{Beutler, F.},
  \bibinfo{author}{Ross, A.~J.} \& \bibinfo{author}{Saito, S.}
\newblock \bibinfo{title}{{Modeling the reconstructed BAO in Fourier space}}.
\newblock \emph{\bibinfo{journal}{Mon. Not. Roy. Astron. Soc.}}
  \textbf{\bibinfo{volume}{460}}, \bibinfo{pages}{2453--2471}
  (\bibinfo{year}{2016}).

\bibitem{Hikage19}
\bibinfo{author}{Hikage, C.}, \bibinfo{author}{Koyama, K.} \&
  \bibinfo{author}{Takahashi, R.}
\newblock \bibinfo{title}{{Perturbation theory for the redshift-space matter
  power spectra after reconstruction}}.
\newblock \emph{\bibinfo{journal}{Phys. Rev. D}}
  \textbf{\bibinfo{volume}{101}}, \bibinfo{pages}{043510}
  (\bibinfo{year}{2020}).


\bibitem{Molino}
\bibinfo{author}{{Hahn}, C.} \& \bibinfo{author}{{Villaescusa-Navarro}, F.}
\newblock \bibinfo{title}{{Constraining M$_{{\ensuremath{\nu}}}$ with the
  bispectrum. Part II. The information content of the galaxy bispectrum
  monopole}}.
\newblock \emph{\bibinfo{journal}{JCAP}} \textbf{\bibinfo{volume}{2021}},
  \bibinfo{pages}{029} (\bibinfo{year}{2021}).

\bibitem{Quijote}
\bibinfo{author}{{Villaescusa-Navarro}, F.} \emph{et~al.}
\newblock \bibinfo{title}{{The Quijote Simulations}}.
\newblock \emph{\bibinfo{journal}{Astrophys. J., Suppl. Ser.}} \textbf{\bibinfo{volume}{250}},
  \bibinfo{pages}{2} (\bibinfo{year}{2020}).

\bibitem{Fisher}
\bibinfo{author}{Tegmark, M.}, \bibinfo{author}{Taylor, A.} \&
  \bibinfo{author}{Heavens, A.}
\newblock \bibinfo{title}{{Karhunen-Loeve eigenvalue problems in cosmology: How
  should we tackle large data sets?}}
\newblock \emph{\bibinfo{journal}{Astrophys. J.}}
  \textbf{\bibinfo{volume}{480}}, \bibinfo{pages}{22} (\bibinfo{year}{1997}).


\bibitem{Zheng07}
\bibinfo{author}{Zheng, Z.}, \bibinfo{author}{Coil, A.~L.} \&
  \bibinfo{author}{Zehavi, I.}
\newblock \bibinfo{title}{{Galaxy Evolution from Halo Occupation Distribution
  Modeling of DEEP2 and SDSS Galaxy Clustering}}.
\newblock \emph{\bibinfo{journal}{Astrophys. J.}}
  \textbf{\bibinfo{volume}{667}}, \bibinfo{pages}{760--779}
  (\bibinfo{year}{2007}).

\bibitem{Sugiyama2024}
\bibinfo{author}{Sugiyama, N.}
\newblock \bibinfo{title}{{Developing a Theoretical Model for the Resummation
  of Infrared Effects in the Post-Reconstruction Power Spectrum}}
  (\bibinfo{year}{2024}).
\newblock \eprint{2402.06142}.

\bibitem{Binfo}
\bibinfo{author}{{Yankelevich}, V.} \& \bibinfo{author}{{Porciani}, C.}
\newblock \bibinfo{title}{{Cosmological information in the redshift-space
  bispectrum}}.
\newblock \emph{\bibinfo{journal}{Mon. Not. Roy. Astron. Soc.}} \textbf{\bibinfo{volume}{483}},
  \bibinfo{pages}{2078--2099} (\bibinfo{year}{2019}).

\bibitem{AP}
\bibinfo{author}{Alcock, C.} \& \bibinfo{author}{Paczynski, B.}
\newblock \bibinfo{title}{{An evolution free test for non-zero cosmological
  constant}}.
\newblock \emph{\bibinfo{journal}{Nature}} \textbf{\bibinfo{volume}{281}},
  \bibinfo{pages}{358--359} (\bibinfo{year}{1979}).

\bibitem{2020JCAP...05..005D}
\bibinfo{author}{{d'Amico}, G.} \emph{et~al.}
\newblock \bibinfo{title}{{The cosmological analysis of the SDSS/BOSS data from
  the Effective Field Theory of Large-Scale Structure}}.
\newblock \emph{\bibinfo{journal}{JCAP}} \textbf{\bibinfo{volume}{2020}},
  \bibinfo{pages}{005} (\bibinfo{year}{2020}).

\bibitem{Samushia:2021ixs}
\bibinfo{author}{Samushia, L.}, \bibinfo{author}{Slepian, Z.} \&
  \bibinfo{author}{Villaescusa-Navarro, F.}
\newblock \bibinfo{title}{{Information content of higher order galaxy
  correlation functions}}.
\newblock \emph{\bibinfo{journal}{Mon. Not. Roy. Astron. Soc.}}
  \textbf{\bibinfo{volume}{505}}, \bibinfo{pages}{628--641}
  (\bibinfo{year}{2021}).



\bibitem{DESI}
\bibinfo{author}{{DESI Collaboration.}} 
\newblock \bibinfo{title}{{The DESI Experiment Part I: Science,Targeting, and Survey Design}}.
\newblock \emph{\bibinfo{journal}{ArXiv e-prints}}  (\bibinfo{year}{2016}).
\newblock \eprint{1611.00036}.

\bibitem{EUCLID}
\bibinfo{author}{Laureijs, R.} \emph{et~al.}
\newblock \bibinfo{title}{{Euclid Definition Study Report}}
  (\bibinfo{year}{2011}).
\newblock \eprint{1110.3193}.

\bibitem{PFS}
\bibinfo{author}{Ellis, R.} \emph{et~al.}
\newblock \bibinfo{title}{{Extragalactic science, cosmology, and Galactic
  archaeology with the Subaru Prime Focus Spectrograph}}.
\newblock \emph{\bibinfo{journal}{Publ. Astron. Soc. Jap.}}
  \textbf{\bibinfo{volume}{66}}, \bibinfo{pages}{R1} (\bibinfo{year}{2014}).

\bibitem{White:2010qd}
\bibinfo{author}{White, M.}
\newblock \bibinfo{title}{{Shot noise and reconstruction of the acoustic peak}}
   (\bibinfo{year}{2010}).
\newblock \eprint{1004.0250}.

\bibitem{2010ApJ...713.1322L}
\bibinfo{author}{{Lawrence}, E.} \emph{et~al.}
\newblock \bibinfo{title}{{The Coyote Universe. III. Simulation Suite and
  Precision Emulator for the Nonlinear Matter Power Spectrum}}.
\newblock \emph{\bibinfo{journal}{Astrophys. J.}} \textbf{\bibinfo{volume}{713}},
  \bibinfo{pages}{1322--1331} (\bibinfo{year}{2010}).

\bibitem{Kobayashi:2020zsw}
\bibinfo{author}{Kobayashi, Y.}, \bibinfo{author}{Nishimichi, T.},
  \bibinfo{author}{Takada, M.}, \bibinfo{author}{Takahashi, R.} \&
  \bibinfo{author}{Osato, K.}
\newblock \bibinfo{title}{{Accurate emulator for the redshift-space power
  spectrum of dark matter halos and its application to galaxy power spectrum}}.
\newblock \emph{\bibinfo{journal}{Phys. Rev. D}}
  \textbf{\bibinfo{volume}{102}}, \bibinfo{pages}{063504}
  (\bibinfo{year}{2020}).


\bibitem{Neveux:2022tuk}
\bibinfo{author}{{Neveux}, R.} \emph{et~al.}
\newblock \bibinfo{title}{{Combined full shape analysis of BOSS galaxies and
  eBOSS quasars using an iterative emulator}}.
\newblock \emph{\bibinfo{journal}{Mon. Not. Roy. Astron. Soc.}}
  \textbf{\bibinfo{volume}{516}}, \bibinfo{pages}{1910--1922}
  (\bibinfo{year}{2022}).
\bibitem{PallEmulator:2023}
\bibinfo{author}{Wang, Y.} \emph{et~al.}
\newblock \bibinfo{title}{{Emulating power spectra for pre- and post-reconstructed galaxy samples}}
  (\bibinfo{year}{2023}).
\newblock \eprint{2311.05848}.


\bibitem{Hahn:2019zob}
\bibinfo{author}{Hahn, C.}, \bibinfo{author}{Villaescusa-Navarro, F.},
  \bibinfo{author}{Castorina, E.} \& \bibinfo{author}{Scoccimarro, R.}
\newblock \bibinfo{title}{{Constraining $M_\nu$ with the bispectrum. Part I.
  Breaking parameter degeneracies}}.
\newblock \emph{\bibinfo{journal}{JCAP}} \textbf{\bibinfo{volume}{03}},
  \bibinfo{pages}{040} (\bibinfo{year}{2020}).
\newblock \eprint{1909.11107}.

\bibitem{Coulton:2022rir}
\bibinfo{author}{Coulton, W.~R.} \emph{et~al.}
\newblock \bibinfo{title}{{Quijote PNG: The information content of the halo
  power spectrum and bispectrum}}  (\bibinfo{year}{2022}).
\newblock \eprint{2206.15450}.

\bibitem{Paillas:2022wob}
\bibinfo{author}{Paillas, E.} \emph{et~al.}
\newblock \bibinfo{title}{{Constraining $\nu \Lambda$CDM with density-split
  clustering}}  (\bibinfo{year}{2022}).
\newblock \eprint{2209.04310}.

\bibitem{Planck2015}
\bibinfo{author}{Ade, P. A.~R.} \emph{et~al.}
\newblock \bibinfo{title}{{Planck 2015 results. XIII. Cosmological
  parameters}}.
\newblock \emph{\bibinfo{journal}{Astron. Astrophys.}}
  \textbf{\bibinfo{volume}{594}}, \bibinfo{pages}{A13} (\bibinfo{year}{2016}).

\bibitem{Tassev:2013pn}
\bibinfo{author}{Tassev, S.}, \bibinfo{author}{Zaldarriaga, M.} \&
  \bibinfo{author}{Eisenstein, D.}
\newblock \bibinfo{title}{{Solving Large Scale Structure in Ten Easy Steps with
  COLA}}.
\newblock \emph{\bibinfo{journal}{JCAP}} \textbf{\bibinfo{volume}{06}},
  \bibinfo{pages}{036} (\bibinfo{year}{2013}).
\newblock \eprint{1301.0322}.

\bibitem{Winther:2017jof}
\bibinfo{author}{Winther, H.~A.}, \bibinfo{author}{Koyama, K.},
  \bibinfo{author}{Manera, M.}, \bibinfo{author}{Wright, B.~S.} \&
  \bibinfo{author}{Zhao, G.-B.}
\newblock \bibinfo{title}{{COLA with scale-dependent growth: applications to
  screened modified gravity models}}.
\newblock \emph{\bibinfo{journal}{JCAP}} \textbf{\bibinfo{volume}{08}},
  \bibinfo{pages}{006} (\bibinfo{year}{2017}).
\newblock \eprint{1703.00879}.

\bibitem{Howlett:2015hfa}
\bibinfo{author}{Howlett, C.}, \bibinfo{author}{Manera, M.} \&
  \bibinfo{author}{Percival, W.~J.}
\newblock \bibinfo{title}{{L-PICOLA: A parallel code for fast dark matter
  simulation}}.
\newblock \emph{\bibinfo{journal}{Astron. Comput.}}
  \textbf{\bibinfo{volume}{12}}, \bibinfo{pages}{109--126}
  (\bibinfo{year}{2015}).
\newblock \eprint{1506.03737}.

\bibitem{Chen19}
\bibinfo{author}{Chen, S.-F.}, \bibinfo{author}{Vlah, Z.} \&
  \bibinfo{author}{White, M.}
\newblock \bibinfo{title}{{The reconstructed power spectrum in the Zeldovich
  approximation}}.
\newblock \emph{\bibinfo{journal}{JCAP}} \textbf{\bibinfo{volume}{09}},
  \bibinfo{pages}{017} (\bibinfo{year}{2019}).

\bibitem{NLrecon}
\bibinfo{author}{Zhu, H.-M.}, \bibinfo{author}{Yu, Y.} \& \bibinfo{author}{Pen,
  U.-L.}
\newblock \bibinfo{title}{{Nonlinear reconstruction of redshift space
  distortions}}.
\newblock \emph{\bibinfo{journal}{Phys. Rev. D}} \textbf{\bibinfo{volume}{97}},
  \bibinfo{pages}{043502} (\bibinfo{year}{2018}).

\bibitem{wrongrecon}
\bibinfo{author}{{Sherwin}, B. D.}, \bibinfo{author}{{White}, M.}
\newblock \bibinfo{title}{{The impact of wrong assumptions in BAO reconstruction}}.
\newblock \emph{\bibinfo{journal}{JCAP}} \textbf{\bibinfo{volume}{02}},
  \bibinfo{pages}{027} (\bibinfo{year}{2019}).

\bibitem{BOSSDR12}
\bibinfo{author}{Beutler, F.} \emph{et~al.}
\newblock \bibinfo{title}{{The clustering of galaxies in the completed SDSS-III Baryon Oscillation Spectroscopic Survey: anisotropic galaxy clustering in Fourier space}}.
\newblock \emph{\bibinfo{journal}{Mon. Not. Roy. Astron. Soc.}}
  \textbf{\bibinfo{volume}{466}}, \bibinfo{pages}{2242--2260}
  (\bibinfo{year}{2017}).

\bibitem{LP1}
\bibinfo{author}{Anselmi, S.}, \bibinfo{author}{Starkman, G.~D.},
  \bibinfo{author}{Corasaniti, P.-S.}, \bibinfo{author}{Sheth, R.~K.} \&
  \bibinfo{author}{Zehavi, I.}
\newblock \bibinfo{title}{Galaxy correlation functions provide a more robust
  cosmological standard ruler}.
\newblock \emph{\bibinfo{journal}{Phys. Rev. Lett.}}
  \textbf{\bibinfo{volume}{121}}, \bibinfo{pages}{021302}
  (\bibinfo{year}{2018}).
  

\bibitem{LP2}
\bibinfo{author}{Anselmi, S.} \emph{et~al.}
\newblock \bibinfo{title}{{Cosmic distance inference from purely geometric BAO
  methods: Linear Point standard ruler and Correlation Function Model
  Fitting}}.
\newblock \emph{\bibinfo{journal}{Phys. Rev. D}} \textbf{\bibinfo{volume}{99}},
  \bibinfo{pages}{123515} (\bibinfo{year}{2019}).

\bibitem{LP3}
\bibinfo{author}{O'Dwyer, M.} \emph{et~al.}
\newblock \bibinfo{title}{{Linear Point and Sound Horizon as Purely Geometric
  standard rulers}}.
\newblock \emph{\bibinfo{journal}{Phys. Rev. D}}
  \textbf{\bibinfo{volume}{101}}, \bibinfo{pages}{083517}
  (\bibinfo{year}{2020}).

\bibitem{LP4}
\bibinfo{author}{Anselmi, S.}, \bibinfo{author}{Starkman, G.~D.} \&
  \bibinfo{author}{Renzi, A.}
\newblock \bibinfo{title}{Cosmological forecasts for future galaxy surveys with
  the linear point standard ruler: Toward consistent bao analyses far from a
  fiducial cosmology}.
\newblock \emph{\bibinfo{journal}{Phys. Rev. D}}
  \textbf{\bibinfo{volume}{107}}, \bibinfo{pages}{123506}
  (\bibinfo{year}{2023}).

\bibitem{Hand17}
\bibinfo{author}{{Hand}, N.}, \bibinfo{author}{{Li}, Y.},
  \bibinfo{author}{{Slepian}, Z.} \& \bibinfo{author}{{Seljak}, U.}
\newblock \bibinfo{title}{{An optimal FFT-based anisotropic power spectrum
  estimator}}.
\newblock \emph{\bibinfo{journal}{JCAP}} \textbf{\bibinfo{volume}{07}},
  \bibinfo{pages}{002} (\bibinfo{year}{2017}).

\bibitem{nbodykit}
\bibinfo{author}{{Hand}, N.} \emph{et~al.}
\newblock \bibinfo{title}{{nbodykit: An Open-source, Massively Parallel Toolkit
  for Large-scale Structure}}.
\newblock \emph{\bibinfo{journal}{Astron. J.}} \textbf{\bibinfo{volume}{156}},
  \bibinfo{pages}{160} (\bibinfo{year}{2018}).

\bibitem{noise}
\bibinfo{author}{{Ando}, S.}, \bibinfo{author}{{Benoit-L{\'e}vy}, A.} \&
  \bibinfo{author}{{Komatsu}, E.}
\newblock \bibinfo{title}{{Angular power spectrum of galaxies in the 2MASS
  Redshift Survey}}.
\newblock \emph{\bibinfo{journal}{Mon. Not. Roy. Astron. Soc.}} \textbf{\bibinfo{volume}{473}},
  \bibinfo{pages}{4318--4325} (\bibinfo{year}{2018}).

\bibitem{Sugiyama:2019ike}
\bibinfo{author}{Sugiyama, N.~S.}, \bibinfo{author}{Saito, S.},
  \bibinfo{author}{Beutler, F.} \& \bibinfo{author}{Seo, H.-J.}
\newblock \bibinfo{title}{{Perturbation theory approach to predict the
  covariance matrices of the galaxy power spectrum and bispectrum in redshift
  space}}.
\newblock \emph{\bibinfo{journal}{Mon. Not. Roy. Astron. Soc.}}
  \textbf{\bibinfo{volume}{497}}, \bibinfo{pages}{1684--1711}
  (\bibinfo{year}{2020}).


\bibitem{Scoccimarro:1999ed}
\bibinfo{author}{Scoccimarro, R.}, \bibinfo{author}{Couchman, H. M.~P.} \&
  \bibinfo{author}{Frieman, J.~A.}
\newblock \bibinfo{title}{{The Bispectrum as a Signature of Gravitational
  Instability in Redshift-Space}}.
\newblock \emph{\bibinfo{journal}{Astrophys. J.}}
  \textbf{\bibinfo{volume}{517}}, \bibinfo{pages}{531--540}
  (\bibinfo{year}{1999}).

\bibitem{Gagrani:2016rfy}
\bibinfo{author}{Gagrani, P.} \& \bibinfo{author}{Samushia, L.}
\newblock \bibinfo{title}{{Information Content of the Angular Multipoles of
  Redshift-Space Galaxy Bispectrum}}.
\newblock \emph{\bibinfo{journal}{Mon. Not. Roy. Astron. Soc.}}
  \textbf{\bibinfo{volume}{467}}, \bibinfo{pages}{928--935}
  (\bibinfo{year}{2017}).

\bibitem{Ballinger:1996cd}
\bibinfo{author}{Ballinger, W.~E.}, \bibinfo{author}{Peacock, J.~A.} \&
  \bibinfo{author}{Heavens, A.~F.}
\newblock \bibinfo{title}{{Measuring the cosmological constant with redshift
  surveys}}.
\newblock \emph{\bibinfo{journal}{Mon. Not. Roy. Astron. Soc.}}
  \textbf{\bibinfo{volume}{282}}, \bibinfo{pages}{877--888}
  (\bibinfo{year}{1996}).

\bibitem{Gil-Marin:2016wya}
\bibinfo{author}{Gil-Mar\'\i{}n, H.} \emph{et~al.}
\newblock \bibinfo{title}{{The clustering of galaxies in the SDSS-III Baryon
  Oscillation Spectroscopic Survey: RSD measurement from the power spectrum and
  bispectrum of the DR12 BOSS galaxies}}.
\newblock \emph{\bibinfo{journal}{Mon. Not. Roy. Astron. Soc.}}
  \textbf{\bibinfo{volume}{465}}, \bibinfo{pages}{1757--1788}
  (\bibinfo{year}{2017}).

\end{thebibliography}

\begin{thebibliography}{30}


\expandafter\ifx\csname url\endcsname\relax
  \def\url#1{\texttt{#1}}\fi
\expandafter\ifx\csname urlprefix\endcsname\relax\def\urlprefix{URL }\fi
\providecommand{\bibinfo}[2]{#2}
\providecommand{\eprint}[2][]{\url{#2}}

\bibitem{2017PhRvD..96d3513H}
\bibinfo{author}{{Hikage}, C.}, \bibinfo{author}{{Koyama}, K.} \&
  \bibinfo{author}{{Heavens}, A.}
\newblock \bibinfo{title}{{Perturbation theory for BAO reconstructed fields:
  One-loop results in the real-space matter density field}}.
\newblock \emph{\bibinfo{journal}{Phys. Rev. D}} \textbf{\bibinfo{volume}{96}},
  \bibinfo{pages}{043513} (\bibinfo{year}{2017}).

\bibitem{Quijote}
\bibinfo{author}{{Villaescusa-Navarro}, F.} \emph{et~al.}
\newblock \bibinfo{title}{{The Quijote Simulations}}.
\newblock \emph{\bibinfo{journal}{Astrophys. J., Suppl. Ser.}} \textbf{\bibinfo{volume}{250}},
  \bibinfo{pages}{2} (\bibinfo{year}{2020}).

\end{thebibliography}
\end{document}